\documentclass[a4paper,12pt]{article}
\usepackage[latin1]{inputenc}
\usepackage[T1]{fontenc}
\usepackage{amsmath,amssymb}
\usepackage{graphicx, color}
\usepackage{theorem,mathrsfs, eufrak}
\usepackage[all]{xy}
\usepackage{bbm,makeidx}
\makeindex

\newtheorem{theo}{Theorem}
\newtheorem{theor}{Theorem}[section]

\newtheorem{defi}{Definition}

\newtheorem{llem}{Lemma}[section]

\newtheorem{conj}{Conjecture}
\newtheorem{prop}{Proposition}[section]
\newtheorem{pprop}{Proposition}

\renewcommand{\Im}{\text{Im}}
\renewcommand{\Re}{\text{Re}}

\newcommand{\der}{\text{d}}
\newcommand{\Is}{\text{Is}}
\newcommand{\Id}{\text{Id}}

\renewcommand{\arg}{\text{Arg}}
\newcommand{\ord}{\text{ord}}

\newcommand{\vol}{\text{vol}}

\newcommand{\tr}{\text{{Tr}}}
\newcommand{\ssup}{\text{singsupp }}
\newcommand{\n}{\text{N}}

\newcommand{\ol}{\overline}

\newcommand{\ds}{\ensuremath{\displaystyle}}
\renewcommand{\cfrac}{\ensuremath{\displaystyle\frac}}

\newcommand{\eps}{\varepsilon}

\newcommand{\K}{\ensuremath{\mathbbm{K}}}

\renewcommand{\L}{\ensuremath{\mathbbm{L}}}
\newcommand{\F}{\ensuremath{\mathbbm{F}}}
\newcommand{\R}{\ensuremath{\mathbbm{R}}}
\newcommand{\PP}{\ensuremath{\mathbbm{P}}}
\newcommand{\T}{\ensuremath{\mathbbm{T}}}
\newcommand{\Q}{\ensuremath{\mathbbm{Q}}}
\newcommand{\C}{\ensuremath{\mathbbm{C}}}
\newcommand{\HH}{\ensuremath{\mathbbm{H}}}
\newcommand{\Z}{\ensuremath{\mathbbm{Z}}}
\newcommand{\N}{\ensuremath{\mathbbm{N}}}
\newcommand{\M}{\ensuremath{\mathbbm{M}}}

\newcommand{\hpl}{\hspace{1em}}
\newcommand{\ghpl}{\hspace{4em}\hspace{4pt}}
\newcommand{\gghpl}{\hspace{6em}\hspace{6pt}}
\newcommand{\hms}{\hspace{-1em}}

\newcommand{\shpl}{\hspace{6pt}}

\newcommand{\vpl}{\vspace{2em}}
\newcommand{\svpl}{\vspace{1em}}
\newcommand{\ssvpl}{\vspace{6pt}}
\newcommand{\sli}{\vspace{3pt}}
\newcommand{\ssli}{\vspace{1pt}}
\newcommand{\vms}{\vspace{-2em}}
\newcommand{\svms}{\vspace{-1em}}
\newcommand{\ssvms}{\vspace{-6pt}}

\newcommand{\np}{\newpage}

\newcommand{\olf}[1]{\overline{#1}^{\F}}

\newcommand{\CC}{\ensuremath{\mathscr{C}}}

\newcommand{\CO}{\ensuremath{\mathscr{O}}}

\newcommand{\CF}{\ensuremath{\mathscr{F}}}
\newcommand{\CL}{\ensuremath{\mathscr{L}}}

\newcommand{\CP}{\ensuremath{\mathscr{P}}}

\newcommand{\CCI}{\ensuremath{\mathscr{I}}}
\newcommand{\CCS}{\ensuremath{\mathscr{S}}}
\newcommand{\CCM}{\ensuremath{\mathscr{M}}}
\newcommand{\CCD}{\ensuremath{\mathscr{D}}}

\newcommand{\CA}{\ensuremath{\EuFrak{A}}}

\newcommand{\CB}{\ensuremath{\EuFrak{B}}}

\newcommand{\CI}{\ensuremath{\EuFrak{I}}}
\newcommand{\CJ}{\ensuremath{\EuFrak{J}}}

\newcommand{\CS}{\ensuremath{\EuFrak{S}}}

\newcommand{\matr}[4]{\displaystyle \left(\begin{array}{cc} #1 & #2
      \\ #3 & #4 \end{array}\right)}
\newcommand{\matrr}[9]{\displaystyle \left(\begin{array}{ccc} #1 & #2 &
      #3 \\ #4 & #5 & #6 \\ #7 & #8 & #9 \end{array}\right)}

\newcommand{\vect}[3]{\displaystyle \left(\begin{array}{c} #1 \\ #2
     \\  #3  \end{array}\right)}
\newcommand{\vvect}[2]{\displaystyle \left(\begin{array}{c} #1 \\ #2
\end{array}\right)}

\renewcommand{\geq}{\geqslant}
\renewcommand{\leq}{\leqslant}

\title{On the scarring of eigenstates for some arithmetic hyperbolic manifolds}

\author{Tristan POULLAOUEC}

\begin{document}

\maketitle
\begin{abstract}
We shall deal here with the conjecture of \emph{Quantum Unique Ergodicity}.
In \cite{R-S}, Rudnick and Sarnak showed that there is no \emph{strong 
scarring} on closed geodesics for compact arithmetic congruence surfaces 
derived from a quaternion division algebra (see Introduction and theorem 
\ref{th1}). 

\sli
We extend this theorem to a class $(K_2^S)$ of Riemannian manifolds $X_{_R}=
\Gamma_{\!_R}\backslash \HH^3$ that are again derived from quaternion division 
algebras, and show that there is no strong scarring on closed geodesics or on 
$\Gamma_{\!_R}$\,-\,closed imbedded totally geodesic surfaces (see theorem \ref{tryst2}).
\end{abstract}


\makeatletter
\@addtoreset{equation}{section}
\@addtoreset{theor}{section}
\@addtoreset{llem}{section}
\@addtoreset{prop}{section}
\makeatother
\def\theequation{\arabic{section}.\arabic{equation}}
\def\thetheor{\arabic{section}.\arabic{theor}}
\def\thellem{\arabic{section}.\arabic{llem}}
\def\theprop{\arabic{section}.\arabic{prop}}
\def\thesection{\arabic{section}}

\def\thedefi{}
\def\thelem{}
\def\thepprop{}
\def\thetheo{}
\def\thecor{}
\def\theconcl{}
\def\theconj{}

\section{Introduction}
\label{intro}
$\bullet$ A topic of great interest in quantum mechanics, and especially in 
quantum chaos, is the study of the limit of the quantized systems when 
$\hbar$\index{hbar@$\hbar$} $\longrightarrow 0$, which is called the 
\emph{semi-classical limit}. The underlying purpose is to identify in the 
quantized system the influence of the classical dynamic (its chaotic nature 
for instance). Let $M$ be a Riemannian manifold of negative sectional curvature 
in dimension two or three, and $\Delta$ the Laplace-Beltrami operator on $M$. 
It is well known (see \cite{Ano}, \cite{Hed} or \cite{Kar}) that the geodesic 
flow on the unitary sphere tangent bundle $T^1 M$ is ergodic and chaotic. After 
quantization, the wave functions of the stationary Schr\"odinger equation 

\ssvms\sli
\begin{equation}
\label{kwa1}
\cfrac{-\hbar^2}{2m} \,\Delta\, \phi+V(q) \phi=E\, \phi
\end{equation}

\sli
\noindent are the $\,\CL^2\,$\index{L2@$\CL^2$} eigenfunctions or \emph{eigenmodes} 
of $\,\frac{-\hbar^2}{2m}\,\Delta+V(q)$. The potential $V(q)$ is in fact related to 
the curvature of $M$. We suppose that this operator has a discrete spectrum 
$(\lambda_k)_{_{k\in\N}}\,$ with $\lambda_k\longrightarrow +\infty\,$ as 
\,$k\longrightarrow+\infty$, which is true at least in the compact case. We denote 
by $\,(\phi_k)_{_{k\in\N}}$ the associated normalized eigenfunctions and by 
$\,(\mu_k)_{_{k\in\N}}$ the corresponding probability measures that are given by

\ssvms\sli
\begin{equation}
\label{kwa2}
\der \mu_k(q)=\big|\phi_k(q)\big|^2 \,\der\vol(q)
\end{equation} 

\sli
\noindent They are actually the probability of presence of a particle in the 
state $\,\phi_k\,$ at $q$, and the semi-classical limit is the limit at large 
energies, that is when $k \longrightarrow +\infty$. The Quantum Unique Ergodicity 
Conjecture (see \cite{Sar}) states :

\begin{conj}
\label{conj1}
Let $M$ be a Riemannian manifold of dimension two or three and of sectional 
curvature $K<0$. Then  $\;\emph{\der} \mu_k \underset{k\rightarrow\infty}
{\longrightarrow} \cfrac{\emph{\der \vol}}{\emph{\vol}(M)}\cdot$
\end{conj}

\sli
For a compact surface $M$ whose geodesic flow is ergodic, this result was 
established in \cite{Zel} for a subsequence of full density of $\left(\mu_k
\right)$. We shall consider more specifically quotient manifolds $M=\Gamma
\backslash\HH^n$ (with $n=2$ or $3$), where $\Gamma$ is a freely acting discrete 
subgroup of $\Is(\HH^n)$. In fact, all Riemann surfaces apart from $S^2$, $\C$, 
$\C^*$ and $\T^2$ are of such a type (see \cite{Far}). A first step towards the 
conjecture was taken in \cite{R-S} with the following result :

\begin{defi}
A probability measure $\nu$ on $M$ is called a \emph{quantum limit} if there 
exists a sequence $\big(\phi_j\big)_{j\in\N}$\index{nu@quantum limit 
\,$\nu$}\index{fi@$\left(\phi_j\right)$, eigenfunctions of $\Delta$} of 
normalized eigenfunctions of $\Delta$ in $\CL^2(M)$ such that the measures 
$\,\left|\phi_j(z) \right|^2 \emph{\der\vol}\,$ converge weakly towards 
$\,\emph{\der}\nu$.
\end{defi}

\begin{theor}
\label{th1}
Let $\,M=\Gamma\backslash \HH^2$ be an arithmetic congruence surface derived 
from a quaternion algebra and $\nu$ a quantum limit on $X$. If $\;\sigma=
\emph{\ssup} \nu\,$ is contained in the union of a finite number of 
isolated points and closed geodesics, then $\,\sigma=\emptyset$.
\end{theor}

\sli
\noindent In other words, there is no \emph{strong scarring} (cf. \cite{Sar}) 
of eigenmodes on closed geodesics of the Riemann surface $M$. In this theorem, 
$\Gamma$ is a congruence subgroup of a discrete group derived from an 
indefinite quaternion division algebra. Actually, we restrict ourselves here to 
\emph{arithmetic quantum limits}, that are quantum limits arising from a sequence 
of joint eigenfunctions of the Laplacian and all Hecke operators. It is expected 
that the spectrum of the Laplacian on $M$ is essentially simple, which justifies 
this restriction. A recent result from \cite{Lin} establishes the Quantum Unique 
Ergodicity Conjecture for compact arithmetic congruence surfaces :

\begin{theor}
\label{th1bis}
Let $\,M=\Gamma\backslash \HH^2$\, with $\Gamma$ a congruence lattice over 
$\Q$. Then for compact $M$ the only arithmetic quantum limit is the 
normalized volume $\der \vol$. For $M$ not compact any arithmetic quantum 
limit is of the form $c\,\der\vol$ with $0\leq c\leq 1$.
\end{theor}

\ssvpl
$\bullet$ We briefly recall some useful notions and results from \cite{R-S} \S 2.1. 

\ssvpl
\index{separates (a correspondence)}
A correspondence \,$\CC$\, of order $\,r$ on a Riemannian manifold $X$ 
is a mapping from\index{CC@$\CC,\,\CC_p$}\index{TC@$T_{_C},\,T_p$} 
$\,X$ to \,$X^r/\CS_r\,$ 
such that $\,\CC(x)=(S_1(x), \dots , S_r(x))$ with $\, S_k \in \Is(X)$ for all
$\,k=1 \dots r$. Here, $\,\CS_r\,$ is the symmetric group of order $\,r$.
\index{IsX@$\Is(X),\,\Is(\HH^n)$} We say that such a correspondence $\,\CC\,$ 
\emph{separates} a subset $\,\Lambda$\, of $X$ if $\,\,\exists\, z\in X-\Lambda\,$ 
such that $\,\exists\,!\, k \in \{1,\dots, r\},\;S_k(z) \in \Lambda$. We shall 
denote by $T_{_C}$ the associated operator of $\CL^2(X)$ defined by $\, T_{_C}(f) 
: x\longmapsto\sum_{k=1}^r f \big(S_k(x)\big)$. 

\begin{prop}
\label{pro1}
Let $\Lambda \subset X$ be a closed subset of zero volume and $\,\CC$ be a 
correspondence on $X$ that separates $\Lambda$. Let $(\phi_j)_{_{j\in\N}}$ 
be a sequence of normalized $\CL^2$\,-\,eigenfunctions of \,$T_{_C}$\, such 
that $\,\emph{\der}\nu=\ds\lim_{j\rightarrow\infty}|\phi_j(z)|^2 \emph{\der 
\vol}(z)\,$ exists. Then $\emph{\ssup} \nu \neq \Lambda$.
\end{prop}

\noindent Keep in mind that the singular support of a probablity measure 
on $X$ is a closed set.

\ssvpl
As we are interested in arithmetic quantum limits, we shall apply this proposition 
to joint eigenfunctions of the Laplacian $\,\Delta$\, and all Hecke operators $\,T_C$.

\section{Complements of hyperbolic geometry and algebra}

\subsection{Hyperbolic geometry in dimension three}\label{GeomHyp}

We take the upper half-space $\,\HH^3=\left\{\,(x,y,t)\in \R^3\;\;\big/\;\; 
t>0\,\right\}$ as a model of hyperbolic space in dimension three. 
\index{H3@$\HH^3$} We set $\,\textbf{j}=1\wedge\textbf{i}\in\R^3\,$ 
and shall identify in the sequel the space $\,\HH^3=\left\{\,z+t\,
\textbf{j}\;\;\big/\;\;z\in\C,\;t>0\,\right\}
$\, with the subset \,$\R\oplus\R\,i\oplus\R_+\textbf{j}\,$ of the algebra of 
quaternions of Hamilton $\HH=\R[1,i,\textbf{j},\textbf{k}]$. This space is 
provided with the Riemannian hyperbolic metric $\,\der s^2=t^{-2} \left(
{\der x}^2+{\der y}^2+{\der t}^2\right)$ having a constant sectional curvature 
$K=-1$. The Laplace-Beltrami operator is $\,\Delta=t^2\left(\frac{\partial^2}
{{\partial x}^2}+\frac{\partial^2}{{\partial y}^2}+\frac{\partial^2}{{\partial 
t}^2}\right)-t \,\frac{\partial}{\partial t}$\, \index{Delta@$\Delta$} and the 
volume form is given by $\der\vol=t^{-3}\, \der x\,\der y\,\der z$ (cf. \cite{Rat} 
\S 4.6).

\begin{figure}[htbp]
  \centering
  \input dess11.pstex_t 
  \caption{ Geodesics of $\HH^3$}
 \label{fig3}
\end{figure}

\ssvpl
The geodesics of $\HH^3$ are the half-circles centered on $\C$ located in 
vertical planes and the half-lines orthogonal to $\C$. They are uniquely 
defined by their ending points in $\,\PP^1(\C)\simeq \C\cup\infty$, two distinct 
points that are the roots of a unique proportionality class of non-degenerate 
complex binary quadratic form : as in dimension two, this association is a bijection 
(cf. \cite{R-S} \S 2.3). Moreover the imbedded totally geodesic submanifolds 
(abbreviated \emph{itgs}) \index{itgs@$\text{itgs}$} of $\,\HH^3$\, are the 
half-planes orthogonal to $\C$ and the half-spheres centered on $\C$. 

\ssvpl
 We know (cf. \cite{Rat} \S 4.4) that $\Is(\HH^3)$\index{IsX@$\Is(X),\,\Is(\HH^n)$}, 
the group of isometries of $\HH^3$, consists of the extensions to $\HH^3$ of the 
M\"obius transformations of $\C$ 

\ssvms\sli
$$\ds \mbox{M}(\C) =  \left\{ \,z \longmapsto \cfrac{az+b}{cz+d}\;, \;
z \longmapsto \cfrac{a\ol{z}+b}{c\ol{z}+d}\;\;\left/\;\; \matr{a}{b}{c}{d} 
\in \mbox{PSL}(2,\C)\,\right.\right\}$$ \index{MC@M$(\C)$}

\ssvms\sli
\noindent and that the subgroup of orientation preserving isometries $\,\Is^+(\HH^3)$\, 
consists of the extensions of the complex fractional linear transformations and is 
isomorphic to PSL$(2,\C)$. We shall make the identification implicitely in the sequel. 
We have the following action of SL$(2,\C)$ on $\HH^3\subset \HH$, called the 
\emph{Poincaré extension}

\ssvms\sli
\begin{equation}
\label{kwa23} 
\forall\,\gamma=\matr{a}{b}{c}{d}\in \mbox{SL}(2,\C)\quad\forall\,x\in\HH^3
\qquad\gamma\cdot x=(ax+b).(cx+d)^{-1}
\end{equation} 

\sli
\noindent whence

\ssvms\sli
$$\forall\,\gamma=\matr{a}{b}{c}{d} \in \mbox{GL}(2,\C)\quad
\mbox{ with }\; ad-bc=n\neq 0\qquad\forall\,(z,t)\in\HH^3$$

 \ssvpl
\noindent if $\,c=0\,$:

\vms\ssvpl
\begin{equation}
\label{kwa24}
\gamma \, . \, \left(\begin{array}{c} z \svpl \\ t \end{array}\right) 
\,=\, \left(\begin{array}{c} \cfrac{a z+b}{d} \ssvpl\sli \\ \left|
\cfrac{a}{d}\right| t\end{array}\right) 
\end{equation}

\sli
\noindent and if $\,c\neq 0\,$:

\svms
\begin{equation}
\label{kwa25}
\gamma\,.\,\left( \begin{array}{c} z \vpl \\ t \end{array} \right) \,=\,
\left(\begin{array}{c} \cfrac{a}{c}-\cfrac{n}{c} \, \cfrac{\ol{c z+d}}{
|c z+d|^2+|c|^2 t^2} \ssvpl\sli \\ \cfrac{|n| t}{|c z + d|^2+|c|^2 t ^2}
\end{array}\right) 
\end{equation}

\ssvpl
The elements of $\Is^+(\HH^3)$ are, exactly like those of $\Is^+(\HH^2)$, 
\index{IsX@$\Is(X),\,\Is(\HH^n)$} characterized by their fixed 
points in $\HH^3\cup \partial \HH^3=\HH^3\cup\C\cup\infty$ (cf. \cite{Rat} \S 4.7). 
Indeed, for $\,ad-bc \neq 0$\, and $\,c\neq 0$

\ssvms\sli
$$\matr{a}{b}{c}{d} \vvect{z}{t}=\vvect{z}{t} \quad \Longleftrightarrow 
\quad \left\{\begin{array}{r@\ c@\ l} a+d=\tr(\gamma)&=&2\,\Re(cz+d) \ssvpl 
\\ |cz+d|^2+|c|^2 t^2&=&1 \end{array}\right.$$
 
\sli
\noindent and for $c=0$

\svms
$$\matr{a}{b}{0}{d} \vvect{z}{t}=\vvect{z}{t} \quad \Longleftrightarrow 
\quad \left\{\begin{array}{r@\ c@\ l} (a-d)z+b&=&1 \ssvpl \\ |a|=|d|&=&1 
\end{array}\right.$$

\sli
\noindent In this case, we have $ad=1$ and, setting $a=r e^{i\theta}$ with 
$r=|a|>0$ and $\theta\in\R$, we get $\tr(\gamma)^2=a^2+2+\frac{1}{a^2}=2+
\left(r^2+\frac{1}{r^2}\right) \cos(2\theta)+i\left(r^2-\frac{1}{r^2}\right) 
\sin(2\theta)$.

\begin{itemize}
\item[$\centerdot$] If $\tr(\gamma)^2\in [0,4[$, the isometry $\gamma$ 
is called \emph{elliptic} (we follow here the classical terminology). 
For $c\neq 0$, we set $\,cz+d=Z=\frac{\tr(\gamma)}{2}+i|c|y$\, with \,$y\in\R$ 
according to the fixed point condition, so that 
	
\svms\sli
$$z=\left(\frac{\tr(\gamma)}{2c}-\frac{d}{c}\right)+i\frac{c}{|c|}\,y
\qquad \mbox{ with }\quad y\in\R$$ 
	
and the forementionned condition writes 
	
\svms
$$4=\tr(\gamma)^2+4\,|c|^2\,(y^2+t^2)$$
	
As a consequence, we have a whole geodesic of fixed points in $\HH^3$, 
the half-circle of radius $\frac{\sqrt{4-\tr(\gamma)^2}}{2|c|}$ centered 
on $\frac{\tr(\gamma)}{2c}-\frac{d}{c}$.\index{ell@elliptic element}
	
\sli
For $c=0$, we have $r=1$ and $\theta\not\equiv 0\,[\pi]$\, since $\tr(
\gamma)^2\in [0,4[$, so that $|a|=|d|=1$ and $a\neq d=\ol{a}$. We obtain 
once again a whole geodesic of fixed points in $\HH^3$, the half-line 
$\left]\frac{1-b}{d-a},\infty\right)$. 
	
\item[$\centerdot$] If $\tr(\gamma)^2=4$ and $\gamma\neq$ I$_2$, the 
isometry $\gamma$ is called \emph{parabolic} and it has a fixed attractive 
point in $\C$, which is $z=\frac{\tr(\gamma)}{2c}-\frac{d}{c}$ for $c\neq 0$ 
and $\infty$ for $c=0$.\index{par@parabolic element}

\item[$\centerdot$] If $\tr(\gamma)^2\notin [0,4[$, the isometry $\gamma$ 
is called \emph{hyperbolic} and it has two fixed points in $\C\cup\infty$, 
one attractive and the other repulsive. The geodesic $L$ connecting these 
points is called the \emph{axis} of $\gamma$, which leaves it invariant and 
acts on it as a translation of the curvilinear abscisse.
\index{hyper@hyperbolic element}
\end{itemize}

\subsection{Algebraic complements}\label{AlgComp}

\subsubsection{Number theory}\label{Numb}

For all notions of number theory, we refer to \cite{Nar} chapter 7, 
\S 2 and \S 3.

\ssvpl
$\bullet$ Let $\K$\index{K@$\K,\,\L$} be a number field and $\L$ a finite 
extension, $\CO_{\K}$ and $\CO_{\L}$ their respective rings of algebraic 
integers. For any prime ideal $\CB$\index{OK@$\CO_{\K},\,\CO_{\L}$}
\index{B@$\CB$} of $\CO_{\L}$, the prime ideal $\,P=\CB\cap\CO_{\K}\,$ of 
$\,\CO_{\K}\,$ is called the \emph{underlying ideal to }$\CB$. Moreover, 
the quantity

\ssvms
 $$\,f_{\L\,/\,\K}(\CB)\overset{def}{=}\big[\,\CO_{\L}/\CB\,:\, 
    \CO_{\K}/P\,\big]\leq\big[\,\L\,:\,\K\,\big] < \infty$$

\index{fLK@$f_{\L\,/\,\K}$} 
\sli
\noindent is called the \emph{degree} of $\CB$ over $\K$. Now take a prime 
ideal $P\,$ of $\,\CO_{\K}$ : we have $\,P \CO_{\L}={\CB_1}^{e_1} \dots 
{\CB_s}^{e_s}$\, with $\CB_i$ a prime ideal of $\CO_{\L}$ of degree $\,f_i= 
f_{\L/\K}(\CB_i)$\, for all $i=1\dots s$. The integers $e_i$ are called 
\emph{ramification indices} and verify

\ssvms
\begin{equation}
\label{kwa21}
\ds \sum_{i=1}^s e_i f_i=\Big[\,\L\,:\,\K\,\Big]
\end{equation}

\noindent If $\, e_i > 1$ for some $i\in \{1,\dots,s\}$, the ideal $P$ is 
\emph{ramified}. There is only a finite number of such ones in $\,\CO_{\K}$.

\svpl
 Let $\,\K\,$ be a number field, \,$P\,$ a prime ideal of $\,\CO_{\K}$, 
\,$p\Z=P \cap \Z\,$ the underlying prime ideal and \,$f=f_{\K/\Q}(P)\,$ 
its degree over \,$\Q$. The \emph{norm} of the prime ideal \,$P$\, is

\ssvms
$$\n(P)\overset{def}{=} |\CO_{\K} / P|= p^f$$

\sli
\noindent A set $A$ of prime ideals of $\CO_{\K}$ is \emph{regular} with 
\emph{density} $a$ in the set of all prime ideals of $\CO_{\K}\,$ if

\svms\ssvms
\begin{equation}
\label{kwa22}
\ds \sum_{P \in A} \n(P)^{-s} \underset{s \rightarrow 1^+}{\sim}
a \log\frac{1}{s-1}
\end{equation}\index{density (of a set of prime ideals)}

\ssvms
\noindent Let $\L/\K$ be a finite extension with normal closure $\M/\K$ 
and Galois group $G=\mbox{Gal}\big(\M/\K\big)$. We know that the set of 
the prime ideals $P$ of $\CO_{\K}\,$ \ssli satisfying $P \CO_{\L}=\CB_1 
\dots \CB_r\,$ with $\CB_i$ a prime ideal of degree $f_{\L/\K}(\CB_i)=f_i$ 
fixed for all $i=1\dots r$ is regular and its density is the relative 
frequence in $G$ of the elements of $G$ that, in the left translation 
representation considered as a~permutation group of the set $G$, are the 
products of $r$ disjoints cycles of length $\,f_1, \dots,\, f_r$ (cf. 
\cite{Nar} proposition 7.15).

\svpl 
Let $\K$ be a number field and take \,$a \in \CO_{\K}\backslash\{1\}\,$ 
square-free. The field $\,\L=\K\left(\sqrt{a}\right)\,$ is a quadratic 
(hence Galois) extension of $\K$, and its Galois group is \,$G=\left\{
\mbox{\Id},\tau\right\}$\, where $\tau^2=\Id$. Moreover $\,\CO_{\L}=
\CO_{\K}[\alpha]\simeq \CO_{\K}[X]/\pi_{\alpha}$ with \,$\alpha=
(1+\sqrt{a})/2\,$ and $\,\pi_{\alpha}=X^2-X+\frac{1-a}{4}\,$ if 
$\,a\equiv 1\,[4]$, \,$\alpha=\sqrt{a}$\, and \,$\pi_{\alpha}=X^2-a$\, 
otherwise. For a prime ideal $P$ of  $\CO_{\K}$, we have $\CO_{\L}/
P\CO_{\L}={\CO_{\K}}_{/(P)}[X]/\pi_{\alpha}$ and $\,P \CO_{\L}={\CB_1}^{e_1} 
\dots {\CB_s}^{e_s}\,$ with $\,\sum_i e_i f_i=\big[\,\L\,:\,\K\,\big]=2$, 
so that only three situations occur :

\begin{itemize}
\ssvms
\item[i)] $P\CO_{\L}=R$, prime in $\CO_{\L}$, is \emph{inert} in iff $a$ is 
not a square modulo $P$ \hpl(\emph{density $1/2$})
\ssvms
\item[ii)] $P\CO_{\L}=R \ol{R}$, with $R$ prime in $\CO_{\L}$, \emph{splits} 
iff $a$ is a square modulo $P$\shpl\,\;(\emph{density $1/2$})
\ssvms
\item[iii)] $P\CO_{\L}=R^2$, with $R$ prime in $\CO_{\L}$, is \emph{ramified} 
iff $a \in P$. \gghpl(\emph{density $0$})
\end{itemize}

\begin{prop}
\label{prop3}
Let $d \in \Z\backslash\{1\}$ be square-free, \,$\K=\Q\big(\sqrt{d}\big)$ a 
quadratic extension and \,$a \in \CO_{\K}$ that is not a square. There exists 
a regular subset $\,\CC\subset\CP$\, of density greater than $1/4$ such that for 
any prime ideal $P$ of $\CO_{\K}$ satisfying $P \cap \Z=p\Z$ with $p\in \CC$, 
$a$ is not a square modulo $P$.
\end{prop}

\sli
\emph{Proof : }let us denote by $A$ the set $\,\left\{ P \mbox{ primes 
of } \CO_{\K}\;/\; a \mbox{ is not a square modulo }P\right\},$ that has  
density $1/2$, and by $B$ the set $\,\left\{\, p\Z=P \cap \Z \;/\; P \in A 
\,\right\}\,$ of underlying ideals of $\Z$. We~shall divide $B$ into three 
subsets $B=B_1 \cup B_2 \cup B_3$, respectively the sets of ideals that are 
inert, ramified or that split. 

\ssvpl
\hpl If $\,P\cap \Z=p\Z \in B_1$ \hpl \hpl\,: \quad $p\CO_{\K}=P \in A$ 
so that $\n(P)=p^2$.

\ssvpl
\hpl If $P\cap\Z=p\Z \in B_2 \cup B_3$ : \quad$p\CO_{\K}=P^2 
\mbox{ or }P\ol{P}$ with $P \in A$, hence $\n(P)=p$.

\ssvpl
\noindent An ideal  $p\Z$ of $B$ either splits ($p\Z \in B_3$) and there are 
two prime ideals of $A$ above it, or it~is inert or ramified ($p\Z \in B_1\cup 
B_2$) and there is only one above it. Therefore

\svms
$$\sum_{P \in A} \n(P)^{-s}= \sum_{p\Z \in B_1} p^{-2s}+  
\sum_{p\Z \in B_2} p^{-s} + 2 \sum_{p\Z \in B_3} p^{-s} \Longrightarrow
 \sum_{P \in A} \n(P)^{-s} \leq 2 \sum_{p\Z \in B} p^{-s}$$

\sli
\noindent Because $A$ has density $1/2$,  $\,B$ contains a subset of density 
greater than $1/4$. 

\svpl 
$\bullet$ Let $\,p \in \CP\,$ and $\,\K\,$ be a field number. Elements 
$\,a_1,\,a_2, \dots,\,a_n\,$ of $\,\K$ are called \emph{$p$-independent} if, 
as soon as $\,{a_1}^{x_1}\,{a_2}^{x_2} \dots {a_n}^{x_n}$ (with $x_i \in \Z$ 
for all $i$) is a $\,p^{th}$ power in \,$\K$, then $\,x_i\equiv 0\,[p]$\, for 
all $\,i=1\dots r$.

\begin{theor}
\label{theo3}
 Let $\,a_1,\dots,\,a_n \in \Z\,$ be $\,2$-independent integers and \,$z_1,
 \dots,\,z_n\in\{\pm1\}$ fixed. There exists infinitely many $p \in \CP$ 
 such that $\,\left(\frac{a_i}{p}\right)=z_i$\, for all $\,i=1\dots n$.
\end{theor}
 
\noindent This follows from the application of theorem 7.13 from \cite{Nar} 
to quadratic characters. 

\begin{prop}
\label{prop4}
 Let $\,a_1,\dots,\,a_n \in \Z \backslash \N$. There exists infinitely many 
 primes $p$ such that $\,\left(\frac{a_i}{p}\right)=-1$\, for all $\,i=1\dots n$.
\end{prop}

\sli
\emph{Proof : }let $a_1,\dots,\,a_n \in \Z \backslash \N$. If they are 
$2$-independent, the result is straight forward by application of theorem 
\ref{theo3}. Otherwise, let us take a maximal $2$-independent subfamily, 
that we can suppose to be $\,a_1,\dots,\,a_m$\, with $ \,1 \leq m <n$, after 
a relabelling. According to theorem \ref{theo3}, there exists infinitely many 
primes $p$ such that

\sli\ssvms
$$\forall \,i=1 \dots m \quad\left(\frac{a_i}{p}\right)=-1$$

\sli
\noindent Let $p$ be such a prime satisfying : $\,\forall \,i=m+1 \dots n,\; 
a_i \not\equiv 0 \,[p]$, and let $\,1 \leq j \leq n-m$. Because the selected 
$2$-independent family is maximal, the elements $\,a_1,\dots$, $\,a_m\,$ and 
$a_{m+j}$ are~not $2$-independent. After simplification

\ssvms
$$\exists\, q \in \Z\quad \exists \,x_1,\dots,\,x_m \in \big\{0,1\big\}
\qquad {a_1}^{x_1} \dots {a_m}^{x_m} a_{m+j}=q^2 > 0$$

\noindent As $a_i <0$ for all $i$, then $\ds \sum_{i=1}^m x_i$ is necessarily 
odd and

\ssvms
$$\hpl\ghpl a_{m+j}=q^2 \prod_{i=1}^m a_i^{-x_i} \;\Longrightarrow\;
\left(\frac{a_{m+j}}{p}\right)=\prod_{i=1}^m
\left(\frac{a_i}{p}\right)^{x_i}=(-1)^{\sum x_i}=-1$$

\sli
\noindent We can take any  \,$j$\, with \,$1\leq j \leq n-m$, which ends the 
proof.

\subsubsection{Quaternion algebras}\label{QuatAlg}

 We first refer to \cite{Eich} \S 1 and \S 2 in this section.

\ssvpl
$\bullet$ Let $a\neq 1$ and $b$ be square-free integers. The 
\emph{quaternion algebra of type $(a,b)$ on $\Q$} is the $\Q$-algebra 
$\CA=\left(\frac{a,b}{\Q}\right)=\Q[1,\omega,\Omega,\omega\Omega]$, 
where $\,\omega^2=a,\,\Omega^2=b$\, and $\,\omega\Omega+\Omega\omega=0$.
\index{A@$\CA$} \index{omega@$\omega,\,\Omega$} The center of $\CA$ is 
$\Q$. Then $\F=\big\{\,q+r\,\omega \;\big/\;q,\;r \in \Q\,\}$ is a 
subfield of $\CA$ identified with $\Q(\sqrt{a})$. We shall write any 
element of $\,\CA\,$ as $\,\alpha=\xi+\eta\,\Omega$\index{xi@$\xi$}
\index{heta@$\eta$} \index{F@$\F$} with $\,\xi,\,\eta=\in\F$. We define 
$\ol{\alpha}=\olf{\xi}-\eta\,\Omega\,$ the \emph{conjugate} of $\alpha$, 
$\tr(\alpha)=\alpha+\ol{\alpha}=\tr(\xi) \in \Q\,$ the \emph{trace} of 
$\alpha$ and $\n(\alpha)=\alpha \ol{\alpha}=\xi\,\olf{\xi}-b\,\eta\,\olf{\eta}
\in\Q\,$ its \emph{norm}.\index{n@$\n$}\index{tr@$\tr$}
Note that $\,\ol{\alpha_1.\alpha_2}=\ol{\alpha_2}.\ol{\alpha_1}$\, for all
$\,\alpha_1,\,\alpha_2 \in\CA$.

\svpl
We call a quaternion algebra \emph{definite} or \emph{indefinite} whether 
its norm is definite ($a<0\,$ and $\,b<0$) or indefinite ($a>0\,$ or $\,b>0$) 
as a quaternary quadratic form on $\R$.\index{def@definite, indefinite}
Keep in mind that $\alpha\in\CA$ is a zero divisor if and only if $\alpha\neq 0$ 
and $\n(\alpha)=0$.

\begin{theo}
$\CA$ has zero divisors $ \;\Longleftrightarrow \;\CA \simeq$ \emph{M}$(2,\Q)$ 
\end{theo}

\noindent In this case, we shall speak of \emph{matrix algebra}. Otherwise, 
we have a \emph{division algebra}\index{A@$\CA$!division/matrix algebra}. In 
any case, the mapping 

\svms
\begin{equation}
\label{kwa27}
\,\varphi \; : \; \alpha=\xi+\eta \,\Omega \longmapsto 
\matr{\xi}{\eta}{b\,\olf{\eta}}{\olf{\xi}}
\end{equation}
\index{ffi@$\varphi$}

\ssvms\sli
\noindent provides the identification of $\CA \otimes \F$ with M$(2,\F)$ and
leaves the trace and the norm (or determinant) invariant. 

\begin{prop}
\label{prop23}
 Let $\,b \in \CP$. If $\;\CA=\left(\frac{a,b}{\Q}\right)\,$ is a matrix algebra, 
then $\,\left(\frac{a}{b}\right)=1$.
\end{prop}

\sli
\emph{Proof : }for a zero divisor $\alpha=x_0+x_1 \omega+x_2 \Omega+x_3 \omega 
\Omega \in \CA$, we have $\alpha\neq 0$ and $\,\n(\alpha)=({x_0}^2-a {x_1}^2)-b 
({x_2}^2-a {x_3}^2)=({x_0}^2- b {x_2}^2)-a ({x_1}^2-b {x_3}^2)=0$. After the 
multiplication by the least common multiple of the denominators of the $x_i$, 
we obtain 

\ssvms
$${y_0}^2- a {y_1}^2 = b \left({y_2}^2-a {y_3}^2\right)\qquad\mbox{where}\quad 
 y_i \in \Z \;\mbox{ for all }\;i=0\dots 3$$
 
\svms
\begin{enumerate}
\item[(i)] If $\,b\,$ does not divide $y_1$ then $a\equiv\left(\frac{y_0}{y_1}
\right)^2\,\big[b\big]$\, is a square modulo $b$.

\ssvms\sli
\item[(ii)] If $\,b\,$ divides $y_1$, this prime divides $\,{y_0}^2\,$ hence 
$\,y_0$ too. By noting $\,y'_0=y_0/b\,$ and $\,y'_1=
y_1/b$, we get

\svms
$${y_2}^2- a {y_3}^2 =b \big({y'_0}^2 - a {y'_1}^2\big)\qquad\mbox{where}
\quad y'_i \in \Z\;\mbox{ for all }\;i=0 \dots 3$$

\ssvms\sli
an equation in $\,y_2$, $\,y_3\,$ of the same type as before. After a finite 
number of simplifications by $\,b$, we are again in case (i), unless $\,y_1=
y_3=0$. In this last case, we have $\,{y_0}^2=b {y_2}^2 \neq 0\,$ since 
$\,\alpha\neq 0\,$, which cannot happen for \,$b\in\CP$. 
\end{enumerate}

\begin{defi} An \emph{order} $\CI$\index{I@$\CI$, $\,\CI_0$, $\,\CI^{pr}$} 
in $\,\CA\,$ is a subring of $\,\CA\,$ such that $1 \in \CI$, $\, \n(\alpha)\,$ 
and $\,\tr(\alpha) \in \Z$\, for all \,$\alpha\in\CI$, and $\,\CI$ has four 
linearly independent generators over $\Q$.
\end{defi}

\noindent Thus, it is a free $\Z$-module of rank four in $\CA$ that is 
besides stable under the conjugation. For example, $\,\CI_0=\CO_{\F}\oplus 
\CO_{\F}\,\Omega\,=\big\{\,\xi+\eta\,\Omega\;\big/\;\xi,\,\eta\in\CO_{\F}\,
\big\}\,$ is an order in the quaternion algebra $\,\CA$, and M$(2,\Z)$ is an 
order in the matrix algebra M$(2,\Q)$. Given $n\in \Z$, we define $\,\CI(n)=
\big\{\,\alpha \in \CI\;\big/\;\n(\alpha)=n\,\big\}\,$ and  $\,\CI^{pr}(n)=
\CI(n)\cap \CI^{pr}\,$ the subset of primitive elements, the ones that cannot 
be divided in $\,\CI\,$ by a non unit integer.

\begin{prop}
\label{prop5}
Let $\CI$ be an order. Then : $\;\exists\, D,\,D' \in \Z \backslash\{0\},\; 
D'\CI \subset D\CI_0 \subset \CI$.
\end{prop}

\sli
\emph{Proof :} since $\,\CI\,$ and $\,\CI_0\,$ are two free $\Z$-modules of 
rank four in $\,\CA$, their $\,\Z$-bases are two $\,\Q$-bases of $\,\CA$. Let 
$\,M\in$ GL$(4,\Q)\,$ be a transition matrix from a basis of $\CI$ to a basis
 of $\CI_0$. Two~integers $\,D\,$ and $\,D'\,$ such that $\,DM \in$ M$(4,\Z)\,$ 
and $\,D'D^{-1} M^{-1} \in$ M$(4,\Z)\,$ satisfy the above property. Moreover, 
$\,\CO_{\F}\simeq \Z^2\,$ and $\CI\simeq \Z^4$ is countable. 

\begin{prop}
\label{prop6}
Let $p \in \CP$ be a prime such that $\ord_p(2abDD')=0$ and $\left(
\frac{a}{p}\right)=-1$. For all $\,\alpha=\xi+\eta \,\Omega \in \CI^{pr}$\, 
such that $\,\n(\alpha) \equiv 0\,[p]$, we have $\,\ord_p{\,\n(\xi)}=\ord_p{
\,\n(\eta)}=0$. In~particular, $\,\xi\,\eta\neq 0$\, for such an $\,\alpha$.
\end{prop}

\sli 
\emph{Proof : }let $p$ and $\alpha$ satisfy the above assumptions ; 
by proposition \ref{prop5}, we have $D'\,\alpha=D\,\alpha_1$ with 
$\alpha_1=\xi_1+\eta_1\,\Omega\in\CI_0$. Besides, $\,\ord_p(D D')=0\,$ 
so that $\,\ord_p{\,\n(\xi)}=\ord_p{\,\n(\xi_1)}\geq 0\,$ and 
$\,\ord_p{\,\n(\eta)}=\ord_p{\,\n(\eta_1)} \geq 0$. Then

\ssvms\sli
$$\n(D'\alpha)=\n\big[D(\xi_1+\eta_1\,\Omega)\big]=
D^2 \big[\n(\xi_1)-b\,\n(\eta_1)\big]\equiv 0\,[p]$$

\sli
\noindent and $\,\n(\xi_1)-b\,\n(\eta_1) \equiv 0 \,[p]$ whence
$\ord_p{\,\n(\xi_1)} > 0  \,\Longleftrightarrow\, \ord_p{\,\n(\eta_1)
}>0$ since $\,\ord_p(b)=0$. Let us assume that $\ord_p{\,\n(\xi)}=
\ord_p{\,\n(\xi_1)} > 0$ and $\ord_p{\,\n(\eta)}=\ord_p{\,\n(\eta_1)}>0$.

\begin{itemize}
\item[\underline{\emph{$a \not\equiv 1 [4]$ }} :] therefore, $\,\CO_{\F}
=\Z[\sqrt{a}]$ and $\,\xi_1=b_0+b_1 \sqrt{a}\,$ with $\,b_0,\,b_1 \in\Z$. 
Relation $\ord_p{\,\n(\xi_1)}>0$ can be expressed as $\,p \,/\,{b_0}^2-a
{b_1}^2$. If $\,p\,/\,b_1,$ then $\,p \,/\, b_0\,$ and $\,\xi_1=p\,\xi_2$\, 
with $\,\xi_2\in \CO_{\F}$. Otherwise $\,a \equiv \left(\frac{b_0}{b_1}
\right)^2[p]\;$ and $\,\left(\frac{a}{p}\right)=1$, a contradiction.

\item[\underline{\emph{$a \equiv 1 [4]$ }} :] in this case, $\,\CO_{\F}=
\Z\left[\frac{1+\sqrt{a}}{2}\right]\,$ and $\,\xi_1=\left(b_0+\frac{b_1}{2}
\right)+\frac{b_1\sqrt{a}}{2}\,$ with $\,b_0,\,b_1 \in \Z$. We have $\,p\,
/\,\n(2\xi_1)\,$ that provides $\, p \,/\, (2b_0+b_1)^2-a {b_1}^2$. If 
$\,p \,/\, b_1,$ then $\, p \,/\, 2b_0+b_1\,$ and $\, p \,/\,b_0$, so that 
$\,\xi_1= p \,\xi_2\,$ with $\,\xi_2 \in \CO_{\F}$. Otherwise, $\,a \equiv 
\left(\frac{2b_0}{b_1}+1\right)^2 [p]\;$ and $\,\left(\frac{a}{p}\right)=1$,
a contradiction.
\end{itemize}

\ssvms
\noindent As a consequence : $\;\exists\,\xi_2 \in \CO_{\F},\;\xi_1=p\,\xi_2$. 
In the same way, we shall show that : $\;\exists\,\eta_2 \in \CO_{\F},\;\eta_1=p\,
\eta_2$. Thus we have $\,D' \alpha=D\big(\xi_1 + \eta_1 \,\Omega \big)= p D\big(
\xi_2+\eta_2\,\Omega\big)=p \,D\alpha_2$\, with $\,\alpha_2\in\CI_0$. 
Because $\,p \wedge D'=1$, there exists $\,x, \,y \in \Z\,$ such that 
$\,xD'+yp=1\,$ whence

\ssvms
\begin{equation}
\label{eqq35}
\alpha=x D' \alpha +p y \alpha=p (x D \alpha_2 + y \alpha)
\end{equation}

\sli
\noindent Since $\,D\alpha_2 \in D \CI_0 \subset \CI$, then $\,\alpha \in 
p\CI$ is not primitive. This contradiction ends the proof.

\svpl
$\bullet$ Let $\CA$ be an indefinite quaternion algebra of type $(a,b)$ 
on $\Q$ (that is $a>0$ or $b>0$). We shall consider in the sequel orders 
of type $(q_1,q_2)$ in $\CA$, which are principal and can be used to define 
modular correspondences (for more details, see \cite{Eich} \S 3). For~$q_1=1$, 
these orders are simply the maximal ones, and we know that each order is 
contained in a maximal one.

\ssvpl
Let $\CI$ be such an order of type $\,(q_1,q_2)$, which we identify 
implicitly with its image $\,R=\varphi(\CI)\,$\index{R@$R,\,R^{pr}$} in 
M$(2,\F)$. We shall also denote by $\,\alpha=\xi+\eta\,\Omega\,$ any element 
of $\,R$. Via Poincar\'e extension, the set $\,R^*=\big\{\alpha\in R\,\big/ 
\,\det(\alpha)\neq 0\big\}\,$ is identified with a subgroup of $\Is^+(\HH^3)$
\index{IsX@$\Is(X),\,\Is(\HH^n)$}. For $n \wedge q_1q_2=1$, we define the 
infinite sets\index{R@$R,\,R^{pr}$}

\svms
$$R(n)=\left\{\alpha\in R\;/\;\n(\alpha)=n\right\}\quad\mbox{and}
\quad R^{pr}(n)=\left\{\alpha\in R\;/\;\alpha\mbox{ primitive and }
\n(\alpha)=n \right\}$$

\sli
\noindent For example, let $\,\CI\,$ be a maximal order containing $\CI_0$ 
and assume that $b>0$ : Pell-Fermat theorem applied to the equation 
$\,{x_0}^2-b{x_2}^2=1\,$ insures that $\,\CI(1)\,$ is infinite.
Let $\,\Gamma_{\!_R}=R(1)\,$\index{GammaR@$\Gamma_{_R}$} be the discrete 
subgroup of SL$(2,\C)\,$ induced by $\,R$. We shall denote $\,X_{_R}=
\Gamma_{\!_R}\backslash \HH^3\,$\index{XR@$X_{_R}$} the quotient space 
and $\,\pi_{\!_R}\;:\;\HH^3\longrightarrow\Gamma_{\!_R}\backslash
\HH^3$\index{pi@$\pi_{_R}$} the canonical projection. To insure 
that $X_{_R}$ inherits  the Riemannian structure of $\HH^3$, the group 
$\,\Gamma_{\!_R}\,$ must not contain any elliptic element (cf. \cite{Rat} \S 8.1).

\begin{prop}  
\label{prop230}
Let us assume that $b \in \CP\backslash\{3\}$. If $\,\Gamma_{\!_R}\,$ 
contains an elliptic element, one of the integers $\,a$, $-a\,$ or $\,-3a$ 
is a square modulo $b$.
\end{prop}\index{ell@elliptic element}

\sli
\emph{Proof : }we shall assume that $\,a\not\equiv 0\,[b]$, otherwise 
$a$ is a square modulo $b$. If \,$\Gamma_{\!_R}$\, contains an elliptic 
element, there exists $\,\alpha=\xi+\eta\,\Omega \in \CI\,$ such that 
$\,|\tr(\alpha)|<2\,$ and $\,\n(\alpha)=1$. Since $\,\CI\,$ is an order, 
$\,\tr(\alpha)=\tr(\xi) \in \big\{-1,0,+1\big\}$. Set $\,\xi=x+y\sqrt{a}\,$ 
and $\,\eta=z+t\sqrt{a} \in \F=\Q[\sqrt{a}]$. Then $\,\n(\alpha)=x^2-a\,y^2-
b(z^2-a\,t^2)=1\,$ and $\,\tr(\alpha)=2x\in\big\{-1,0,+1\big\}$. 

\begin{itemize}
\item[\underline{\emph{$x=0$}} :] thus $\,-a y^2=1+b (z^2-at^2)\,$ 
and after multiplication by the least common multiple of the denominators 
of $y$, $z$ and $t$, we get $\,-a {y'}^2=E^2+b ({z'}^2-a{t'}^2)\,$ with 
integers $y'$, $z'$ and $t'$ such that $\,y'\wedge z'\wedge t'=1$. 
If $\,y'\not\equiv 0\,[b]$, $\,-a\equiv E^2/{y'}^2\,[b]\,$ is a square 
modulo $b$. If $\,y'\equiv 0\,[b]$, $\,b$ divides $E\,$ and $\,{z'}^2-
a{t'}^2\equiv 0\,[b]\,$ whence $\,t'\not\equiv 0\,[b]\,$  because 
$\,y'\wedge z'\wedge t'=1\,$. Therefore $\,a\equiv {z'}^2/{t'}^2\,[b]\,$ 
is a square modulo $b$.

\item[\underline{\emph{$x=\pm 1/2$}} : ] then $\,-a y^2=3/4+b(z^2-
at^2)\,$ and after multiplication by the least common multiple of the 
denominators of $y$, $z$ and $t$, we get $\,-a {y'}^2=3 E^2+b ({z'}^2-
a{t'}^2)\,$ with integers $y'$, $z'$ and $t'$ such that $\,y'\wedge 
z'\wedge t'=1$. Proceeding as in the case $\,x=0$, we deduce that one of 
the integers $-3a$ or $a$ is a square modulo $b$, which ends the proof 
of the proposition.
\end{itemize}

\ssvpl
For all $\,n\in\N\,$ such that $\,n\wedge q_1 q_2=1$,\, the set $R(1)
\backslash R(n)=\left\{\Gamma_{\!_R}\, \alpha_1,\dots,\, \Gamma_{\!_R} 
\,\alpha_r \right\}$ is finite (cf. \cite{Eich} \S 7) and we may define 
the \emph{modular correspondance} of order $n$ on the quotient space 
$X_{_R}=\Gamma_{\!_R} \backslash \HH^3$ by

\svms
$$\ds \CC_n : \left| \begin{array}{ccl} X_{\!_R} & \longrightarrow & 
{X_{\!_R}}^r / \CS_r \ssli \\ \Gamma_{\!_R} \,x & \longmapsto & 
\{\Gamma_{\!_R} \,\alpha_1\, x, \dots, \, \Gamma_{\!_R} \,\alpha_r\, 
x\}\end{array}\right.$$

\sli
\noindent The associated modulor operators $T_n$ of $\CL^2(X_{_R})$ 
are defined by

\sli\ssvms
\begin{equation} 
\label{eqq170}
\forall\, f\in \CL^2(X_{_R})\quad \forall\,z\in\HH^3\qquad
T_n\,(f)(\Gamma_{\!_R} z)=\ds \sum_{\alpha \in R(1)\backslash R(n)} 
f(\Gamma_{\!_R}\, \alpha\, z)
\end{equation}

\noindent They are bounded linear operators on $\CL^2(X_{\!_R})$ and
satisfy the classical properties of the modular operators (self 
adjointness, commutation with the Laplace-Beltrami operator $\Delta$, 
composition ...). We shall only consider in the sequel modular 
correspondences $\CC_p$ and modular operators $\,T_p\,$ for $\,p\in
\CP\setminus\{2\}$.\index{CC@$\CC,\,\CC_p$}\index{TC@$T_{_C},\,T_p$}

\begin{prop}
\label{prop13}
 For $\,p\in\CP\,$ such that $p\wedge q_1q_2=1$, we set 
$|R(1) \backslash R(p)|=m$ and denote $R(1) \backslash 
R(p)=\big\{\Gamma_{\!_R}\,\alpha_1,\dots, \,\Gamma_{\!_R}\,\alpha_m
\big\}$. Then  we have : $\forall\,i,\exists\,!\,j $ such that 
$\,\alpha_j\,\alpha_i \in p\,R(1)$\, and $\,\forall \,k \neq j,\;
\alpha_k\,\alpha_i \in R^{pr}(p^2).$
\end{prop}

\sli
\emph{Proof : }for all $\,n \in \N$, as $\CI$ is stable under the conjugation, 
the set $R(n)$ is stable under the passage to the transpose of the comatrix, 
which corresponds to the inversion in $\Is^+(\HH^3)\simeq \mbox{PSL}(2,\C)$

\svms\ssvms\sli
$$\varphi(\xi+\eta\,\Omega)=\matr{\xi}{\eta}{b\olf{\eta}}{\olf{\xi}} \in 
R(n)\Longrightarrow \varphi(\ol{\xi+\eta\,\Omega})=\matr{\olf{\xi}}{-\eta}
{-b\olf{\eta}}{\xi} \in R(n)$$

\ssvms\sli
\noindent Take $i\in\{1,\dots,m\}$ : there is $\,j\in \{1,\dots,m\}\,$ 
such that $\,\Gamma_{\!_R}\,{}^t$Com$(
\alpha_i)=\Gamma_{\!_R}\,\alpha_j$.  Therefore $\,\Gamma_{\!_R}\,\alpha_j
\,\alpha_i=p\,\Gamma_{\!_R}\,$ and $\,\Gamma_{\!_R}\,\alpha_k\,\alpha_i\neq 
\Gamma_{\!_R}\,\alpha_j\,\alpha_i=p\,\Gamma_{\!_R}\,$ for all $\,k\neq j$.

\ssvpl
\emph{Remark : }the self-adjointness of the modular operators can be 
quite easily deduced from this proposition.

\section{The quotient space $X_{_R}=\Gamma_{\!_R}\backslash\HH^3$}
\label{QuotSpace}

\subsection{The class of $\big(K_2\big)$\,-\,manifolds}
\index{A@$\CA$!division/matrix algebra}\label{zebuth}

In an indefinite division algebra $\CA=\left(\frac{a,b}{\Q}\right)$ 
with $a<0$ and $b>0$ square-free integers, we consider an order 
\,$\CI$\, of type $(q_1,q_2)$ such that $\,\Gamma_{\!_R}\,$ acts 
freely on $\,\HH^3$. 
This way, we get a class $\,\big(K_2\big)$\, of Riemannian manifolds 
$\,X_{_R}=\Gamma_{\!_R}\backslash\HH^3$\, that have a sectional curvature 
\,$K=-1$, when provided with the metric induced by $\,\HH^3$. This class 
$\,\big(K_2\big)\,$ is far from being empty. Indeed, take $\,a\in\Z_-
\backslash\{-1,-3\}\,$ square-free and $\,b\in\CP\backslash\{3\}\,$ such 
that $a$ is not a square modulo $b$\, : \,$\CA$ is an indefinite division 
quaternion algebra according to proposition \ref{prop23}. For the action 
of $\,\Gamma_{\!_R}\,$ to be free, we just have to impose that $-1$ and 
$-3$ are squares modulo $b$, by proposition \ref{prop230}. Given $a$ 
fixed, theorem \ref{theo3} shows that these three conditions modulo $b$ 
are simultaneously satisfied by infinitely many primes $b$, because the 
negative integers $\,a$, $-1$ and $-3\,$ are 2-independent.

\begin{defi}
The \emph{class $\big(K_2^S\big)$} is the infinite set of 
$\big(K_2\big)$\,-\,manifolds such that

\ssvms\sli
\begin{equation}
\label{eqq181}
a\in\Z_-,\;b\in\CP,\;\left(\cfrac{a}{b}\right)=-1,\;\left(
\cfrac{-1}{b}\right)=1,\;\left(\cfrac{-3}{b}\right)=1
\end{equation}
\end{defi}

\index{K2S@class $(K_1^S)$, class $(K_2^S)$}
\noindent For example, we can take $\,a=-2$, $\,b=13$\, and $\CI$ a 
maximal order containing $\CI_0=\CO_{\F}\oplus\CO_{\F}\,\Omega$. In the 
sequel of this work, we shall consider manifolds $X_{_R}$ of the class 
$\big(K_2^S\big)$. The conjugation in $\F\simeq\Q\big(\sqrt{a}\big)$ 
coincides with the complex conjugation because $a<0$.
 
 \sli
\begin{llem}
\label{lem6}
Let $\,\xi \in \F\backslash\{0\}$. Then $\,\ord_b\, |\xi|^2\,$ is even.
\end{llem}

\sli
\emph{Proof : }let $\xi\,$ be such an element, that we will write 
$\,\xi=p q^{-1}\,\big(x+y\,\sqrt{a}\big){q}$\, with $q,\,p,\,x,\,y\in\Z$,  
$\,x \wedge y =1\,$ and $\,p \wedge q=1$. Assume that$\,|x+y\,\sqrt{a}|^2
=x^2-a\,y^2\equiv 0\,[b]$. If $b$ divides $y$, then $b$ divides $x$, which 
contradicts $x\wedge y=1$. Therefore $\,y \not \equiv 0 \,[b]$ so that
\,$a \equiv \left(\frac{x}{y}\right)^2 \,[b]$\, and $\,\left(\frac{a}{b}
\right)=1$, a contradiction with relation (\ref{eqq181}). Hence $\,|x+y\,
\sqrt{a}|^2 \not\equiv 0 \,[b]\,$ and, as $\,|\xi|^2=p^2\,q^{-2}\,|x+y\,
\sqrt{a}|^2$, we deduce that $\,\ord_b\,|\xi|^2=2\,\big[\ord_b\,(p)
-\ord_b\,(q)\big]\,$ is even. 

\sli
\begin{prop}
\label{prop8}
Let $X_{_R}$ be a $\big(K_2^S\big)$\,-\,manifold. Then $\,\Gamma_{\!_R}\,$ 
has no parabolic element.
\end{prop}\index{par@parabolic element}

\emph{Proof : }let $\,\gamma=\xi+\eta\,\Omega\,$ in $\,\Gamma_{\!_R}$\, be 
parabolic. By taking its opposite $-\gamma$ if necessary, we may assume 
that $\,\xi= 1+ x \sqrt{a}\,$ and $\,\eta=y+z\sqrt{a}\,$ with $\,x,\,y\,$ 
and $\,z\in\Q$. We have then $\,\n(\gamma)-1=0=|\xi|^2-b\,|\eta|^2-1=-a x^2
-b\,(y^2-a z^2)$. Let us multiply \,$x$, $y$ and $z\,$ by the least 
common multiple of their denominators and divide the obtained integers by 
their greatest common divisor. We get $\,a X^2+b\,(Y^2- a Z^2)=0$\, with 
$\,X,\,Y,\,Z \in \Z$\, and $\,X\wedge Y \wedge Z =1$. Because $\,b \wedge 
a =1$, $\,b$ divides $X$. Setting $X_0=X/b$, we get after simplification 
$\,ab{X_0}^2+Y^2-a Z^2=0$. If \,$b$\, divides \,$Z$, $\,b$\, divides $\,Y$\, 
too, a contradiction with relation $\,X \wedge Y \wedge Z =1$. Therefore 
$\,Z \not \equiv 0 \, [b]\,$ and $\,a\equiv\left(\frac{Y}{Z}\right)^2\,[b]$\, 
whence $\,\left(\frac{a}{b}\right)=1$, a contradiction with relation 
(\ref{eqq181}).

\sli
\begin{prop}
\label{prop25}
Let $\CL$ be a closed geodesic of  $X_{_R}=\Gamma_{\!_R}\backslash \HH^3$. 
There exists an hyperbolic transformation $\,\gamma\in\Gamma_{\!_R}\,$ 
whose axis $\,L\subset \HH^3\,$ projects onto $\,\CL\,$ in $\,X_{_R}$.
\end{prop}

\noindent The proof (see \cite{Rat} \S 9.6 for a general form) uses only the 
discontinuity of the action of $\,\Gamma_{\!_R}\,$ on $\,\HH^3$. 
In particular, there exists a compact segment $\,l\,$ of the geodesic 
$\,L\,$  such that $\,L=\bigcup_{n\in\Z} \gamma^n\cdot\, l\,$ and 
$\,\CL=\pi_{\!_R}(L)=\pi_{\!_R}(l)$. Using this Proposition, we show :
\index{hyper@hyperbolic element}

\begin{llem}
\label{lem10}
Let $\CL_1$ and $\CL_2$ be two closed geodesics of $X_{_R}=\Gamma_{\!_R}
\backslash \HH^3$. Then $\,\CL_1=\CL_2\,$ or $\,\CL_1 \cap \CL_2\,$ is 
finite.\index{closed@closed}
\end{llem}

\sli
\emph{Proof : }let $\,\gamma_1\,$ and $\,\gamma_2 \in \Gamma_{\!_R}\,$ be 
hyperbolic transformations whose axis $\,L_1\,$ and $\,L_2\,$ project 
onto $\,\CL_1\,$ and $\,\CL_2\,$ respectively. Let $\,l_1\,$ and $\,l_2\,$ 
be compact segments of $\,L_1\,$ and $\,L_2\,$ such that $\,L_1=\bigcup_{n
\in\Z}\,{\gamma_1}^n\cdot l_1\,$ and $\,L_2=\bigcup_{n\in\Z}\,{\gamma_2}^n
\cdot l_2$. Then 

\ssvms
$$\CL_1 \cap \CL_2
=\pi_{\!_R}\left[\Big(\ds \cup_{_{\gamma\,\in\,\Gamma_{\!_R}}}\gamma 
\cdot l_1\Big) \cap\Big(\ds\cup_{_{\gamma'\,\in\,\Gamma_{\!_R}}} 
\gamma' \cdot l_2\Big) \right]=\pi_{\!_R}\left[\ds\cup_{_{\gamma\,
\in\,\Gamma_{\!_R}}} \Big( l_1 \cap  \gamma\cdot l_2\Big)\right]$$

\sli
\noindent Assume that $\CL_1\cap \CL_2$ is infinite. The group 
$\Gamma_{\!_R}$ acts discontinuously on $\HH^3$ and $\,l_1\,$ and 
$\,l_2\,$ are compact subsets, so that $\,\Gamma_0=\big\{\;\gamma\in
\Gamma_{\!_R}\;\big/\;l_1\cap\gamma\cdot l_2 \neq\emptyset\;\big\}\,$ 
is finite. Since $\CL_1\cap \CL_2=\pi_{\!_R}\left(l_1 \cap  \Gamma_0\cdot 
l_2\right)$ is infinite, there exists $\,\gamma\in\Gamma_0\,$ such that 
$\, l_1\cap\gamma\cdot l_2\,$ hence $\,L_1 \cap\gamma\cdot L_2\,$ is 
infinite. $\,L_1$ and $\, \gamma\cdot L_2\,$ being two geodesics of 
$\,\HH^3\,$ \emph{i.e.} half-circles or half-lines, we deduce that 
$\,L_1= \gamma\cdot L_2\,$ and $\,\CL_1=\CL_2$. 

\subsection{Imbedded surfaces in $\,\big(K_2^S\big)$\,-\,manifolds}

The itgs of $\,X_{\!_R}$\, are quite simply the images under 
$\,\pi_{\!_R}$\, of those of $\,\HH³$, that are the half-spheres 
centered on $\,\C\,$ and the half-planes orthogonal to $\,\C$. 
\index{itgs@$\mbox{itgs}$}\index{trace (of an itgs)}

\subsubsection{Traces of itgs}

\begin{defi}
Let $\,\CCS\,$ be an itgs of $\,\HH³$. Its \emph{trace $\,\CC\,$ on 
$\,\C\,$} is the set of its limit points in $\,\C$, that is $\,\CC=
\ol{\CCS} \cap \C$.
\end{defi}

\noindent The trace of an itgs of $\,\HH³\,$ is then either a circle 
or straight line of $\,\C$ -- \emph{i.e.} a circle of $\,\PP^1(\C)$. 
Moreover, each itgs its uniquely defined by its trace on $\,\C$, whence 

\begin{pprop}
There is a bijection between the itgs of $\,\HH^3\,$ and the circles of 
$\,\PP^1(\C)$.
\end{pprop} 

\noindent As a consequence, the action of an isometry on an itgs in 
$\,\HH^3\,$ is entirely determined by the former's action on the latter's 
trace in \,$\C$, which is much more easy to deal with. In particular 

\begin{prop}
\label{brune}
 Let $\,\CCS_1\,$ and $\,\CCS_2\,$ be two itgs of $\,\HH^3$\, of 
 traces $\,\CC_1\,$ and $\,\CC_2$. Then 

\ssvms
$$\forall \,\gamma \in \mbox{\emph{SL}}(2,\C) \qquad \gamma\cdot
\CCS_1=\CCS_2 \;\Longleftrightarrow \;\gamma(\CC_1)=\CC_2$$
\end{prop} 

\subsubsection{About closed itgs}

The \emph{closed} itgs of $\,X_{\!_R}$\, are the compact ones. The 
lifting to $\HH^3$ of a closed itgs of $\,X_{\!_R}$\, is called 
\emph{closed for $\,\Gamma_{\!_R}$}. Then :

\begin{prop}
\label{prop7}
Let $\,\CCS\,$ be a closed itgs of $\,X_{_R}\,$ and $\,S\,$ be a lifting 
to $\,\HH^3$. There exists a group $\,\Gamma_0\subset\Gamma_{\!_R}\,$ and 
a compact subset $\,\CF \subset S\,$ with non zero area such that 

$$\gamma \in \Gamma_0 \;\Longleftrightarrow\;\gamma\cdot S =S\quad
\mbox{ and }\quad S=\Gamma_0\cdot\CF=\ds \bigcup_{\gamma \in \Gamma_0}
\gamma\cdot\CF$$

\noindent There exists moreover $\,\gamma=\xi+\eta\, \Omega \in 
\Gamma_0\,$ hyperbolic.
\end{prop}

\ssvpl
\emph{Proof : }the projection $\,\CCS=\pi_{_R}(S)$\, being compact, 
there exists $\,\CF \subset S\,$ compact and $\,\Gamma' \subset 
\Gamma_{\!_R}$\, such that $\,S=\Gamma'\cdot \CF$. Since $\,\Gamma_{\!_R}$\, 
is countable, the set $\,\CF\,$ has non zero area (in $\,S$). \ssli Let us 
set $\,\Gamma_0=\left\{\,\gamma\in\Gamma_{\!_R}\;\;\big/\;\;\gamma\cdot\CF
\subset S\,\right\}\,\supset \Gamma'$. For $\,\gamma\in\Gamma_0$, the set 
$\,\gamma\cdot S \cap S \supset \gamma\cdot \CF$\, has non zero area. As 
the itgs $\,\gamma\cdot S\,$ and $\,S\,$ are half-planes or half-spheres, 
$\,\gamma\cdot S =S$. Conversely, the relation $\,\gamma\cdot S=S\,$ implies 
that $\,\gamma\cdot\CF\subset\gamma\cdot S =S\,$ and $\,\gamma\in\Gamma_0$. 
Hence, \ssli $\,\Gamma_0=\big\{\,\gamma\in\Gamma_{\!_R}\;/\;\gamma\cdot S=
S\,\big\}$\, is obviously a group. From the definition of the class 
$\big(K_2^S\big)$ and proposition \ref{prop8}, we know that $\,\Gamma_{\!_R}\,$ 
contains only hyperbolic elements except for \,$\{\pm\Id\}$. Now $\,\Gamma_0
\neq \{\pm\Id\}\,$ otherwise \,$S=\CF$\, would be a compact itgs in $\HH^3$, 
and we can find a hyperbolic element $\,\gamma=\xi+\eta\,\Omega\in\Gamma_0$. 
Then $\,b\,|\eta|^2=|\xi|^2-\n(\gamma)=|\xi|^2-1\geq\Re^2(\xi)-1>0$.

\sli
\begin{llem}
\label{lem666}
Let $ \,\CCS_1\,$ and $\,\CCS_2\,$ be two distinct closed itgs of 
$\,X_{_R}$. Their intersection $\,\CCS_1 \cap \CCS_2\,$ is 
 either the empty set or a closed geodesic of $\,X_{_R}$.
\end{llem}

\sli
\emph{Proof : }in a Riemannian manifold, two distinct itgs intersect 
transversally, because an itgs is entirely defined by a point and the 
tangent space at this point. Then their intersection has dimension one 
if it is not empty. Since $\,\CCS_1\,$ and \,$\CCS_2$\, are closed itgs 
of $\,X_{_R}$, they are compact. Hence, $\,\CL=\CCS_1\cap \CCS_2\,$ is 
a compact subset of $\,X_{_R}$. If $\,\CL\neq\emptyset$, it is a complete 
geodesic because for \,$(M,\vec{u}) \in T\CL$, the geodesic of \,$X_{_R}\,$ 
tangent to $\,\vec{u}$\, at $\,M$\, is contained in both \,$\CCS_1$\, and 
\,$\CCS_2$. As \,$\CL$\, is compact, it is then a closed geodesic of $\,X_{_R}$.

\svpl
The half-sphere $\,S^{o}=S\big(\mbox{O},1/\sqrt{b}\big)$\, is invariant 
under the action of all the isometries induced by $\,\CA \otimes \R$ : 
indeed, given $\,\gamma=\xi+\eta\,\Omega\in \CA \otimes \R\,$ such that
 \,$\n(\gamma)=|\xi|^2-b\,|\eta|^2 \neq 0$, we have\index{S0@$S^{o}$}

\svms\ssvms
\begin{equation}
\label{eqqbienbourrable}
\forall\,\theta \in \R\qquad \big|\gamma(b^{-1/2}\,e^{i\theta})\big|=
\left|\cfrac{\xi e^{i\theta}b^{-1/2}+\eta}{b\,\ol{\eta}e^{i\theta}b^{-1/2}+
\ol{\xi}}\right|=\cfrac{1}{\sqrt{b}}\left|\,\cfrac{\xi\,e^{i\theta}+ 
\eta\,\sqrt{b}}{\ol{\eta}\,\sqrt{b}+\ol{\xi}\,e^{-i\theta}}\right|=
\cfrac{1}{\sqrt{b}}
\end{equation}
 
\noindent whence $\,\gamma\cdot\left(S^{o} \cap \C\right)=S^{o}\cap \C\,$ 
and $\,\gamma\cdot S^{o}=S^{o}$. We also denote by $\,S^{o}$\, its 
projection in $\,X_{\!_R}$. Unfortunately, as we shall see in Appendix 
\ref{stgiferm}, it is the only closed itgs of $\,X_{\!_R}$\, : 
indeed, the subgroup $\,\Gamma_0\subset\Gamma_{\!_R}\,$ leaving 
an itgs $\,S\,$ invariant is generically a one-parameter group (cf. 
$\CI$ is a modulus of rank four) so that $\Gamma_0\backslash S$ cannot 
be compact.

\subsubsection{Notion of $\,\Gamma_{\!_R}$\,-\,closed itgs}

Lacking of closed itgs in our space $X_{_R}$, we shall use instead 
the follwing weaker notion, defined by analogy with the closed geodesics 
of $\,X_{\!_R}$, whose liftings to $\,\HH^3$\, are invariant under a 
hyperbolic element of $\,\Gamma_{\!_R}$
\index{GammaR@$\Gamma_{_R}$!$\Gamma_{_R}$\,-\,closed} :

\begin{defi}
 An itgs \,$S$\, of \,$\HH^3$\, is called \emph{$\,\Gamma_{\!_R}$\,-\,closed} 
if there exists $\,\gamma=\xi+\eta\,\Omega\in\Gamma_{\!_R}\,$ hyperbolic
 such that $\,\gamma\cdot S=S$. Its projection $\,\CCS\,$ in $\,X_{\!_R}\,$ 
is also called \emph{$\,\Gamma_{\!_R}$\,-\,closed}.
\end{defi}

\sli
\noindent The closed itgs are $\,\Gamma_{\!_R}$\,-\,closed, but the converse 
is false. Indeed, we shall show in Appendix \ref{stgigammaferm} the existence
 of infinitely many $\,\Gamma_{\!_R}$\,-\,closed itgs in $\,\HH^3$, then
 verify that their projections define infinitely many distinct itgs of 
$\,X_{_R}$. Therefore, there are infinitely many $\,\Gamma_{\!_R}$\,-\,closed 
itgs in $\,X_{_R}$. 

\begin{llem}
\label{llem666a}
Let $\,\CCS_1\,$ and $\,\CCS_2\,$ be to distinct $\,\Gamma_{\!_R}$\,-\,closed 
itgs of $\,X_{_R}$. Then we have area$\,(\CCS_1 \cap \CCS_2)=0$.
\end{llem}

\sli
\emph{Proof : }it is a straight forward corollay of the first part of 
the proof of lemma \ref{lem666} since every one-dimensional set has zero 
area. Just be aware that the notion of area is here inherent to the 
manifold $\,\CCS_1$\,  provided with the Riemannian metric induced by 
$\,\HH^3$\, : we~can indeed only mention the area of subsets of 
two-dimensional imbedded manifolds of $\,X_{\!_R}$. 

\begin{defi}
 We say that $\,\Lambda \subset X_{_R}$\, is of \emph{type $(S^{o})$} if 
$\,\Lambda$\, is contained in a finite union of $\,\Gamma_{\!_R}$\,-\,closed 
itgs and \,area$\,(\Lambda)=$ area$\,(\Lambda \cap S^{o}) \neq 0$.
\index{S0@$S^{o}$!type \,$S^{o}$}
\end{defi}

\subsection{Isometries and itgs of $\,\HH^3$}\label{isoitgs}

For the sequel, we shall need to exploit relations such as 
$\,\gamma\cdot \CCS=\CCS'$\, for an isometry $\,\gamma$\, and 
two itgs $\,\CCS$\, and $\,\CCS'$\, of $\,\HH³$, at least to 
characterize  $\Gamma_{\!_R}$\,-\,closed itgs. All the half-planes 
and the half-spheres considered in the sequel are implicitely itgs
of $\,\HH^3$.
 
\subsubsection{Of the half-planes}\label{SepPlans}

\begin{prop}
\label{zyva}
Let $\,\CP\,$ be a half-plane, $\,\CCD$\, its trace and $\,\gamma=
\matr{a}{b}{c}{d}\in$ \emph{SL}$(2,\C)\,$ with $\,a+d \neq 0\,$ 
and $\,c\neq 0$. Then 

\ssvms
$$\ghpl\gamma\cdot \CP=\CP \;\;\Longleftrightarrow\;\;\left\{\; 
\begin{array}{l} \frac{a}{c}\,\in\,\CCD,\;\frac{-d}{c}\,\in\,\CCD\;
:\;\,\CCD\,\mbox{ is given by relation }(\ref{eqq102}) \hms\hms\hms\hms
\hms\sli\sli  \\ (a+d)^2 \in \R \end{array} \right.$$
\end{prop}

\sli
\emph{Proof : }by proposition \ref{brune}, we have \,$\gamma\cdot\CP=\CP 
\;\Longleftrightarrow \; \gamma(\CCD)=\CCD$. If $\,\gamma(\CCD)=\CCD,$ 
then $\,\gamma(\infty)=\frac{a}{c} \in \CCD\,$ and $\,\gamma^{-1}(\infty)
=-\frac{d}{c} \neq \frac{a}{c}\in\CCD$. As a consequence,

$$z \in \CCD \; \Longleftrightarrow \; \frac{z-\gamma^{-1}(
\infty)}{\gamma(\infty)-\gamma^{-1}(\infty)}=\cfrac{cz+d}{a+d}
\in\R \;\Longleftrightarrow\;\cfrac{cz-a}{a+d} \in \R  $$

\noindent whence 

\svms
\begin{equation}
\label{eqq102}
\CCD=\left\{\; z \in \C\quad\Big/\quad \Im\left(\cfrac{cz}{a+d}
\right)=\Im\left(\cfrac{a}{a+d}\right)\;\right\}
\end{equation}

\ssvpl
\noindent Moreover $\,\frac{a}{c}\in\CCD=\gamma(\CCD)\,$ so that 

\ssvms\sli
$$\begin{array}{rcl} \gamma\left(\cfrac{a}{c}\right)= \cfrac{a^2+bc}{c(a+d)} 
\in \CCD & \Longleftrightarrow & \cfrac{a^2+bc-a(a+d)}{(a+d)^2}\in \R 
\ssvpl \\ & \Longleftrightarrow & \left(\cfrac{bc-ad}{(a+d)^2}\right)
\in\R \ssvpl\sli \\ & \Longleftrightarrow & (a+d)^2 \in \R\end{array} $$

\sli
\noindent Reciprocally, assume that $\,\frac{a}{c}=\gamma(\infty) \in 
\CCD$, $-\frac{d}{c}=\gamma^{-1}(\infty) \in \CCD\,$ and $\,(a+d)^2 
\in \R$ : then relation (\ref{eqq102}) still holds and $\,\gamma(
\frac{a}{c}) \in \CCD$. Moreover $\,\gamma\left(\frac{a}{c}\right)= 
\frac{a^2+bc}{c(a+d)} \neq \frac{a}{c}\,$ since $\,bc=ad-1 \neq ad$.
Therefore, the isometry $\,\gamma\,$ takes the three distinct points 
$\,\infty$ , $\,\frac{a}{c}$, $\,-\frac{d}{c}\,$ of $\,\CCD\,$ into 
$\,\frac{a}{c}$, $\,\gamma(\frac{a}{c})\,$ and $\,\infty$, which also 
are distinct points of $\,\CCD$ : any circle of $\,\PP^1(\C)\,$ being 
uniquely defined by three points, we have indeed $\,\gamma(\CCD)=\CCD$, 
which ends the proof.

\svpl
Keep in mind that for any hyperbolic element $\,\gamma=\xi+\eta\,\Omega
\in  \Gamma_{\!_R}$, we have $\,\tr(\gamma)=\tr(\xi) \neq 0\,$ and $\,\eta 
\neq 0$ (cf. proof of proposition \ref{prop7}) : those elements, which are 
the only interesting ones for us, will satisfy the hypothesis of the above 
proposition.

\subsubsection{Of the half-spheres}

\begin{prop}
\label{oboidorman}
 Let the half-spheres $\,\CCS_1=S(a_1,r_1)\,$ and $\,\CCS_2=S(a_2,r_2)\,$ 
be itgs of $\,\HH³$, \,$N\in\Z\,$ and $\,\alpha=\xi+\eta\,\Omega\in 
R^{pr}(N)\,$ with $\,\eta\neq 0$. Then $\,\alpha(\CCS_1)=\CCS_2$\, iff

 \ssvms
$$\qquad \begin{array}{c} \exists\, \eps=\pm 1 \\ \mbox{ } \\ \mbox{ }
\end{array}\qquad\left\{\begin{array}{r@\ c@\ lcl} b\big(r_1\,\ol{\eta}
\,a_2-\eps\,r_2\,\eta\,\ol{a_1}\big)&=&\big(r_1+\eps r_2\big)\xi&\quad 
&(\ref{eqq49})\sli\\ b^2\,{r_1}^2\,|\eta|^2-|\xi+b\,\eta\,\ol{a_1}|^2 & 
= & N\eps\,\cfrac{r_1}{r_2} & & (\ref{eqq48})\sli\\ |\xi|^2-b|\eta|^2 & 
= &N & & (\ref{eqq47})\end{array} \right.$$
\end{prop}
 
\ssvpl
\emph{Proof : }by proposition \ref{brune}, we have $\,\alpha(\CCS_1)=
\CCS_2\iff\alpha(\CC_1)=\alpha(\CC_2)$, where $\,\CC_1=C(a_1,r_1)$\, 
and $\,\CC_2=C(a_2,r_2)$\, are the traces of $\,\CC_1$\, and $\,\CCS_2$.
Moreover,

\ssvms\sli
$$\forall\,z\in\C,\quad\alpha(z)= \cfrac{\xi z+\eta}{b\ol{\eta} z +
\ol{\xi}}=\cfrac{\xi}{b \ol{\eta}}+\cfrac{b |\eta|^2-|\xi|^2}{
b\ol{\eta}\left(b\ol{\eta} z+\ol{\xi}\right)}=\cfrac{\xi}{b\ol{\eta}} 
-\cfrac{N}{b\ol{\eta}\left(b\ol{\eta} z+\ol{\xi}\right)}$$

\noindent so that

\svms
 \begin{equation}
\label{eqq39}
\forall\,z \in \C, \quad\alpha(z)=\cfrac{\xi}{b \ol{\eta} }+
\cfrac{k}{z-\zeta}\qquad\mbox{where }\;\;k=-\cfrac{N}{b^2\ol{\eta}^2}
\;\;\mbox{ and }\;\;\zeta=-\ol{\xi}/b\ol{\eta}
\end{equation}

\begin{figure}[htbp]
  \centering
  \input dess30.pstex_t 
\caption{ Action of $\alpha$ on $\CC_1$}
  \label{fig1666}
\end{figure}

\ssvpl
\emph{\underline{Direct implication :} }assume that $\,\alpha(\CC_1)=
\CC_2$. Note that  $\,\zeta\notin \CC_1\,$ otherwise $\,\alpha( \zeta)=
\infty \in \CC_2$, a contradiction. Relation (\ref{eqq39}) implies that 
$\,\alpha=\alpha_R\circ \alpha_I,$ where $\,\alpha_I\,:\, z \longmapsto 
\zeta+|k|/(\ol{z}-\ol{\zeta})\,$ is an inversion of center $\,\zeta\,$ 
and $\,\alpha_R\,$ is an~orientation reversing euclidean isometry of $\C$.

\sli\sli
First we assume that $\,\zeta\neq a_1$. Let Inv$_{\CC_1}$ be the inversion 
of circle $\,\CC_1$\, and $\,\hat{\zeta}=$Inv$_{\CC_1}(\zeta)=a_1+{r_1}^2/(
\ol{\zeta}-\ol{a_1})$. All the circles passing through $\,\zeta\,$ and 
$\,\hat{\zeta}\,$ are orthogonal to $\,\CC_1$ because they are invariant 
under Inv$_{\CC_1}$. Let $\,\CC_0\,$ be such a circle : $\,\alpha_I(\zeta)=
\infty$\, so that $\,\alpha_I\cdot\CC_0\,$ is a line orthogonal to $\,
\alpha_I\cdot\CC_1=\CC'$, hence a diameter. Moreover it contains the point 
$\alpha_I(\hat{\zeta})$. As a consequence $\,\alpha_I(\hat{\zeta})$, the 
intersection of all the diameters of $\,\CC'$, is the center $a'$ of 
$\,\CC'$\, and $\,\alpha(\hat{\zeta})=\alpha_R\circ\alpha_I(\hat{\zeta})=
a_2\,$ is the center of $\,\CC_2$. We set $\,\big(\zeta,a_1\big)\cap 
\CC_1=\big\{A_1,B_1\big\}$\, and $\,\big(\zeta,a_1\big)\cap\CC'=\big\{
A',B'\big\}$, taking here as a convention that the points $\,A_1\,$ and 
$\,\zeta\,$ are on the same side of $\,a_1\,$ on the~line $\,(\zeta,a_1)$\, 
whereas $\,B_1\,$ and $\,\zeta\,$ are on opposite sides. We have therefore 
$\,A'=\alpha_I(B_1)$, $\,B'=\alpha_I(A_1)\,$ and 

\ssvms
$$A_1-\zeta=\cfrac{a_1-\zeta}{|a_1-\zeta|}\,\big(|a_1-\zeta|-r_1\big),
\quad B_1-\zeta=\cfrac{a_1-\zeta}{|a_1-\zeta|}\,\big(|a_1-\zeta|+r_1\big)$$

\noindent whence

\svms
$$A'=\zeta+\cfrac{|a_1-\zeta|}{\ol{a_1}-\ol{\zeta}}\, \cfrac{|k|}{|a_1
-\zeta|+r_1}\hpl\mbox{and}\hpl B'=\zeta+\cfrac{|a_1-\zeta|}{\ol{a_1}-
\ol{\zeta}}\, \cfrac{|k|}{|a_1-\zeta|-r_1}$$

\sli
\noindent Therefore $\,2 r_2=|A'-B'|=2 |k| r_1 / \big||a_1-\zeta|^2-{
r_1}^2\big|$. We set $\,\eps=1\,$ if $\,\zeta\,$ is inside of $\,\CC_1\,$ 
and $\,\eps=-1\,$ otherwise, so that 

\ssvms
\begin{equation}
\label{eqq46}
\cfrac{r_2}{r_1}=\cfrac{\eps |k|}{{r_1}^2-|a_1-\zeta|^2}
\end{equation}

\sli
\noindent As $\,\hat{\zeta}-\zeta=a_1-\zeta+{r_1}^2/\big(\ol{\zeta}-
\ol{a_1}\big)=\big({r_1}^2-|a_1-\zeta|^2\big)/\big(\ol{\zeta}-\ol{a_1}
\big)=\eps |k| r_1/r_2 \big(\ol{\zeta}-\ol{a_1}\big)$, we deduce from 
relation (\ref{eqq39}) that $\,\alpha(\hat{\zeta})=\xi/b\ol{\eta}+\eps
\big(\ol{\zeta}-\ol{a_1}\big)\,k\,r_2/|k|\,r_1$. Besides, $\,|k|=N/b^2
|\eta|^2\,$ so that $\,k/|k|=-|\eta|^2/\ol{\eta}^2=-\eta/\ol{\eta}\,$ and 

$$\,\alpha(\hat{\zeta})=\xi/b\ol{\eta}-\eps\,\eta\,r_2\big(\ol{\zeta}-
\ol{a_1}\big)/\,\ol{\eta}\,r_1=a_2$$ 

\noindent This is equivalent to

\svms
\begin{equation}
\label{eqq49}
b\big(r_1\ol{\eta} a_2- \eps\,r_2\eta\ol{a_1}\big)=\big(r_1+\eps r_2
\big) \xi\end{equation}

\sli
\noindent Injecting $\,\zeta=-\ol{\xi}/b\ol{\eta}\,$ and $|\,k|=N/b^2|
\eta|^2\,$ into the relation (\ref{eqq46}), we get

\ssvms\sli
\begin{equation}
\label{eqq48}
b^2 {r_1}^2 |\eta|^2-|\xi+b\eta\ol{a_1}|^2=N\eps\,\cfrac{r_1}{r_2}
\end{equation}

\ssvms\sli
\noindent At last, $\,\n(\alpha)=N$\, so that 

\svms\ssvms
\begin{equation}
\label{eqq47}
|\xi|^2-b|\eta|^2=N
\end{equation}

\sli\sli
In the case $\zeta=a_1=-\ol{\xi}/b\ol{\eta}$, we have $\,\eps=1$, 
$\,\hat{\zeta}=\infty\,$ and $\,\alpha(\hat{\zeta})=\xi/b\ol{\eta}=a_2\,$ 
so that the~relation (\ref{eqq49}) is still satisfied. Moreover, $\,\CC'=
C(a_1,r_2)$ : $\,\alpha_I\cdot C(a_1,r_1)=C(a_1,r_2)\,$ whence $\,|k|=
r_1 r_2=N/b^2|\eta|^2\,$ and relation (\ref{eqq48}) holds. We still have 
relation (\ref{eqq47}).

\svpl
\emph{\underline{Backward implication :} }assume that the three relations 
hold and keep the previous notations. We deduce from relation (\ref{eqq48}) 
that 

\ssvms\sli
$$|\zeta-a_1|²-{r_1}²=|\ol{\xi}/b\,\ol{\eta}+a_1|^2-{r_1}^2 \neq 0$$
 
\sli
\noindent whence $\,\zeta\notin \CC_1$. Therefore, $\,\alpha_I(\CC_1)=
\CC'\,$ is a circle and $\,\alpha(\CC_1)=\alpha_R(\CC')=\CC''=C(a'',r'')$. 
We still have, by relation (\ref{eqq48}), $\,\eps=1\,$ iff $\,\zeta\,$ is 
inside of $\,\CC_1$. Applying the direct implication to $\,\CC_1\,$ and 
$\,\CC''$, we deduce from relation (\ref{eqq46})

$$\cfrac{r''}{r_1}=\cfrac{\eps |k|}{{r_1}^2-|a_1-\zeta|^2}=\cfrac{\eps\,N}
{b^2\,{r_1}^2\,|\eta|^2 - |\xi+b\,\eta\,\ol{a_1}|^2}=\cfrac{r_2}{r_1}$$

\sli
\noindent because $|k|=N/b^2\,|\eta|^2$. Then $\,r''=r_2$. Finally, 
$\,\alpha_I(\hat{\zeta})\,$ is the~center of $\,\CC'$ so that $\,\alpha(
\hat{\zeta})=a''$ ; as relation (\ref{eqq49}) is equivalent to $\,\alpha(
\hat{\zeta})=a_2$, then $\,a''=a_2$\, whence $\,\alpha(\CC_1)=\CC_2$. 

\ssvpl\sli
Now we can characterize the invariance of a half-sphere under a hyperbolic
 element. The following result applies in particular to closed itgs of 
$\,X_{\!_R}$.

\begin{prop}
\label{pro7}
Let $\,S(a_1,r)\neq S^{o}\,$ be a $\,\Gamma_{\!_R}$\,-\,closed itgs of 
$\,\HH^3$ : therefore $\,a_1\neq 0.$ If \,$q=1+b\,(|a_1|^2-r^2)\neq 0$, 
then $\,\zeta=\frac{a_1}{q} \in \F^*\,$ and $\;\exists\,(X,Y)\in\Z\times
\Q\,$ such that \,$a\,(1-4b|\zeta|^2)=(X^2-4) \,Y^2 > 0$.
\end{prop}

\emph{Proof : }let us take a  hyperbolic element $\,\gamma=\xi+\eta\,
\Omega\in\Gamma_{\!_R}\,$ such that $\,\gamma(\CC)=\CC=C(a_1,r)$. If 
$\,a_1=0$, then $\,\gamma\cdot C(0,r)=C(0,r)\,$ \emph{ i.e.} $\,\big|
\gamma(r\,e^{i\theta})\big|=r$\, for all $\,\theta\in \R$. Thus

\ssvms\sli
\begin{equation}
\label{eqq119}
\forall\,\theta\in\R\qquad\big|\xi\,r\,e^{i\theta}+\eta\big|=r\,
\big|b\,\ol{\eta}\,r\,e^{i\theta}+\ol{\xi}\big|=\big|\xi\,r\,e^{i\theta}+
b\,r^2\,\eta\big|
\end{equation}

\sli 
\noindent By taking the maxima of both sides, we get $\,r\,|\xi|
+|\eta|=r\,|\xi|+b\,r^2\,|\eta|$. As $\,\gamma\,$ is hyperbolic, 
$\,\eta \neq 0\,$ and $\,r=1/\sqrt{b}\,$ whence $\,\CCS=S^{o}$, 
a~contradiction. Therefore $\,a_1\neq 0$. We shall now assume that 
$\,q=1+b\,(|a_1|^2-r^2)\neq 0$. By proposition \ref{oboidorman}, 
the relation $\,\gamma(\CC) =\CC\,$ leads to 

\ssvms
$$\qquad \begin{array}{c} \exists\, \eps=\pm 1 \\ \mbox{ } \\ \mbox{ } 
\end{array} \qquad \left\{\begin{array}{r@\ c@\ lcl} b\big(\ol{\eta}\, 
a_1- \eps\,\eta\,\ol{a_1}\big)& = &\big(1+\eps \big)\, \xi & \quad & 
(\ref{eqq49}') \sli \\ b^2\, r^2 \,|\eta|^2-|\xi+b\,\eta\,\ol{a_1}|^2& = 
&\eps & & (\ref{eqq48}')\sli \\ |\xi|^2-b|\eta|^2 & = &1 & & (\ref{eqq47}')
\end{array} \right.$$

\sli
\noindent If $\,\eps=1\,$, we deduce from (\ref{eqq49}') that \,$\Re(\xi)=0$, 
and this contradicts the hyperbolicity of $\,\gamma$. Therefore 
$\,\eps=-1$, and relation $(\ref{eqq49}')$ implies that $\,\ol{\eta}
\,a_1+\eta\,\ol{a_1}=0$. Summing  relations (\ref{eqq48}') and 
(\ref{eqq47}'), we get 

\ssvms
$$2\,b\,\Re(\xi\,\ol{\eta}\,a_1)+b^2\,(1+|a_1|^2-r^2)\,|\eta|^2=0$$ 

\sli
\noindent As $\,\eta\neq 0$\, and $\,2\,b\,\Re(\xi\,\ol{\eta}\,a_1)=
b\,\big(\xi\,\ol{\eta}\,a_1+\ol{\xi}\,\eta\,\ol{a_1}\big)=b\,\big(\xi-
\ol{\xi}\big)\,\ol{\eta}\,a_1$, we have 

\sli\ssvms
$$\,\big(\xi-\ol{\xi}\big)\,a_1+\big[\underbrace{
1+b\,(|a_1|^2-r^2)}_{q} \big] \,\eta =0$$

\noindent Indeed $\,\xi-\ol{\xi}\neq 0\,$ and $\,\zeta$, equal to 
$\,a_1/q= \eta/(\ol{\xi}-\xi)\in \F^*$, satisfies $\,\eta=(\ol{\xi}
-\xi)\,\zeta= -2 i\Im(\xi)\,\zeta$. Injecting this into the relation 
(\ref{eqq47}'), we obtain 

\ssvms
$$1=\Re(\xi)^2+(1-4\,b\,|\zeta|^2)\,\Im(\xi)^2$$ 

\sli
\noindent Finally setting finally $\,2\,\Re(\xi)=X \in \Z\,$ and 
$\,2\,\Im(\xi)/\sqrt{-a}=Y^{-1}\in\Q^*$, we get after multiplication 

\ssvms
$$\,4=X^2- a\,(1-4\,b\,|\zeta|^2)\,Y^{-2}\,$$

\noindent whence 

\ssvms
$$\exists\,(X,\, Y)\in \Z \times \Q \qquad 
a\,(1-4\,b\,|\zeta|^2)=Y^2\,(X^2-4) > 0$$

\sli
\noindent because $\,Y \neq 0\,$ and $\,X^2=\tr^2(\gamma) > 4\,$ as 
$\,\gamma\,$ is hyperbolic. This ends the proof.

\section{Separation results in $\big(K_2^S\big)$-manifolds}
\label{Statement}

We shall prove the following extension of theorem \ref{th1} to the 
$\big(K_2^S\big)$-manifolds.

\begin{theor}
\label{tryst2}
Let $\,X_{_R}=\Gamma_{\!_R} \backslash \HH^3$ be a three-dimensional 
manifold, $\,\Gamma_{\!_R}$ being a discrete subgroup of $\,\Is^+(
\HH^3)$\, derived from an indefinite quaternion algebra $\,\CA=\left(
\frac{a,\,b}{\Q}\right)$. We shall assume moreover that $X_{_R}$ is a 
manifold of class $\big(K_2^S\big)$.

\ssvpl
Let $\,\Lambda\subset X_{_R}\,$ be a non empty set contained in a finite 
union of $\,\Gamma_{\!_R}$\,-\,closed itgs of $X_{_R}$ such that 
area$(\Lambda)\neq0$, unless $\Lambda$ is contained in a finite union of 
isolated points and closed geodesics, and that is not of type $(S^{o})$.
Then $\,\Lambda\,$ cannot be the singular support of an arithmetic quantum 
limit on $\,X_{_R}$.
\end{theor}

\subsection{Foreword}

First we prove a separation result on such a subset $\Lambda$ :

\begin{prop}
\label{prop15} 
Let $X_{_R}$ be a manifold of class $\left(K_2^S\right)$. For all 
non-empty subset $\Lambda \subset X_{_R}$ contained in a finite union 
of $\,\Gamma_{\!_R}$\,-\,closed itgs such that area$(\Lambda)\neq0$, 
unless $\Lambda$ is contained in a finite union of isolated points and 
closed geodesics, and that is not of type $(S^{o})$, there exists a 
correspondence $\CC$ separating $\Lambda$.
\end{prop}
\index{Lambda@$\Lambda$}

\sli
We shall consider a subset $\Lambda$ of a finite union of 
\emph{objects of the same type} of $\,X_{_R}\,$ (a set of points, a set 
of closed geodesics or a set of $\,\Gamma_{\!_R}$\,-\,closed itgs) and 
treat the cases separately in the next sections. The following proposition 
will simplify the calculations and be very helpful in the sequel.

\begin{prop}
\label{prop15666}
Let $\,F_1,\,\dots,\,F_r\,$ be \;\emph{objects of the same type} of 
$\,X_{_R}\,$ and \,$G_1,\,\dots,\,G_r\,$ a choice of liftings of these 
objects in $\HH^3$. There exists a finite subset $\CF\subset \CP\,$ 
such that, given $\,p\in\CP\backslash\CF$, the~relation  

\ssvms
$$\exists\, \alpha \in R(p)\cup R^{pr}(p^2) \quad
\exists\, i\in\{1 \dots r\}\qquad \alpha\cdot G_1=G_i$$

\ssvms
\noindent leads to

\svms
\begin{equation}
\label{decadix}
\exists\, N \in \CF \quad \exists\, \tilde{\alpha} \in R^{pr}(N\,p)
\cup R^{pr}(N^2p^2) \qquad \tilde{\alpha}\cdot G_1=G_1
\end{equation}
\end{prop}
                        
\ssvpl
\emph{Proof : }let us fix $\,n=1\,$ or $\,2$ and assume that 
\,$\exists\,p_i\in\CP$, $\,\exists\,\alpha_i \in R^{pr}({p_i}^n)$\, 
such that $\,\alpha_i\cdot G_1=G_i$\, whence $\,G_1={ }^t 
\mbox{Com}(\alpha_i)\cdot G_i$, for a certain $\,i\in\{1\dots r\}$. 
Take $\,p\neq p_i \in \CP\,$ and $\,\alpha \in R^{pr}(p^n)\,$ : 
then $\,\alpha\cdot G_1=G_i\, \Longrightarrow\,\tilde{\alpha}\cdot 
G_1=G_1\,$ where \,$\tilde{\alpha}={ }^t\mbox{Com}(\alpha_i)\,\alpha 
\in R({p_i}^n p^n)=R(N^n p^n)$. This element is primitive : otherwise, 
we would have $\,\tilde{\alpha} \in pR\,$ or $\,\tilde{\alpha}\in 
p_i R\,$ since $p$ and $p_i$ are both primes, so that\index{Com@Com}

\begin{enumerate}

\ssvms\sli
\item[.] if $\,\tilde{\alpha} \in pR\,$ we get $\,\alpha_i\,\tilde{
\alpha}=\alpha_i\, { }^t\mbox{Com}(\alpha_i)\,\alpha={p_i}^n \alpha \in 
pR\,$ and $\,\alpha \in pR$\, as $p_i\wedge p=1$, a~contradiction with 
$\,\alpha\in R^{pr}(p^n)$.

\ssvms\sli
\item[.] if $\,\tilde{\alpha} \in p_i R,$ then $\,\tilde{\alpha}\,
{ }^t\mbox{Com}(\alpha)={ }^t\mbox{Com}(\alpha_i)\,\alpha\,{ }^t\mbox{
Com}(\alpha)=p^n { }^t\mbox{Com}(\alpha_i) \in p_i R\,$ and 
$\,\alpha_i \in p_i R$, a~similar contradiction.

\ssvms\sli
\end{enumerate}

\noindent Proceeding the same way with all the indices $\,i\in\{
1\dots r\}\,$ and all the values of $n\in\{1,2\}$, we get to relation 
(\ref{decadix}) after exclusion of at most $2r$ values of $\,p\in\CP$, 
the forementionned set $\,\CF$.

\subsection{Case of the points}\label{SepPoints}

In this section, we only consider primes $p$\, such that $\ord_p(2abDD')=0$ 
and $\,\left(\frac{a}{p}\right)=-1$.

\ssvpl
$\bullet$ Let $\,\Lambda=\big\{\tilde{x}_1,\dots,\,\tilde{x}_l \big\}\,$ 
be a set of points of  $\,X_{_R}$\, and their liftings $\,x_i=(z_i,t_i) 
\in \HH^3\,$ for $\,i=1\dots l$, to which we apply proposition 
\ref{prop15666}. For $\,n=1\,$ or $ \,2$, \,$N\in\CF\,$ and $\,p\in\CP 
\backslash\CF$, we take $\,\alpha=\xi+\eta\,\Omega \in R^{pr}(N^n p^n)\,$ 
such that $\,\alpha\cdot x_1=x_1$. By proposition $\ref{prop6}$, we have 
$\,\xi\,\eta\neq 0$. By relation (\ref{kwa25}), the action of $\,\alpha$\, 
on \,$\HH^3$\, is 

\ssvms\sli
$$\alpha\cdot \vvect{z}{t}=
\vvect{ \cfrac{\xi}{b\ol{\eta}} - \cfrac{N^n p^n}{b\ol{\eta}}\,
\cfrac{\xi+b\eta \ol{z} }{|\xi+b\eta \ol{z}|^2+b^2 |\eta|^2 t^2 }
\ssvpl}{\cfrac{N^n p^n t}{|\xi+b\eta \ol{z}|^2+b^2 |\eta|^2 t^2 }}$$

\sli
\noindent It is well defined for any $t>0$ because $\eta\neq 0$. 
The relation $\,\alpha\cdot x_1=x_1$\, implies that

\begin{equation}
\label{eqq103}
 N^n p^n=|\xi+b\,\eta\,\ol{z_1}|^2+b^2\,|\eta|^2
\,{t_1}^2= |\xi|^2-b\,|\eta|^2
\end{equation}

\ssvms
\noindent and

\svms\ssvms
\begin{equation}
\label{eqq104}
\ol{\eta}\,z_1+\eta\,\ol{z_1}=0
\end{equation}

\sli
\noindent Note that $\,z_1\neq 0$, otherwise  relation (\ref{eqq103}) 
gives $\,|\xi|^2+b^2\,|\eta|^2\,{t_1}^2=|\xi|^2-b\,|\eta|^2\,$ so that 
$|\eta|=0$, a contradiction. Relation (\ref{eqq104}) implies that 
$\,-\frac{\ol{z_1}}{z_1} =\frac{\ol{\eta}}{\eta}$\, is a constant. Choose 
a $\,\eta_0 \in \CO_{\F}\,$ such that $\,\frac{\ol{\eta_0}}{\eta_0}=
-\frac{\ol{z_1}}{z_1}$. We have $\,\frac{\ol{\eta}}{\eta}=\frac{\ol{\eta_0}
}{\eta_0}$\, whence \,$\frac{\eta}{\eta_0}=\frac{\ol{\eta}}{\ol{\eta_0}}
\in\R\cap\F=\Q\,$ (because \,$a<0$) and $\,\exists \, m \in \Q\,$ such 
that $\,\eta=m\,\eta_0$. The expansion of relation (\ref{eqq103}) provides

\svms
$$N^n\,p^n=|\xi|^2+2\,b\,m\,\Re\,(\xi_0\,\ol{\eta_0}\,z_1)+b^2\,m^2
\,|\eta_0|^2\,({t_1}^2+|z_1|^2)=|\xi|^2-b\,m^2\,|\eta_0|^2\,$$

\noindent and after division by $\,b\,m \neq 0$\, we get

\ssvms
\begin{equation}
\label{eqq106}
2\,\Re\,(\xi\,\ol{\eta_0}\,z_1)+m\,|\eta_0|^2\,
\big[1+b\,({t_1}^2+|z_1|^2)\big]=0
\end{equation}

\sli
\noindent As $\,D'\,\xi \in \CO_{\F}$ by proposition \ref{prop5}, we 
have $\,2\,D'\xi=X+Y\sqrt{a}\,$ with $\,X,\,Y\in\Z$. Because $b>0$, the 
coefficient of $\,m\,$ in relation (\ref{eqq106}) is strictly positive 
so that $\,m\,$ is a linear function of $\,X\,$ and $\,Y$. Therefore 
the middle term in (\ref{eqq103}) is a definite positive quadratic form 
of the two integer variables $\,X\,$ and $\,Y$, that we will write

\ssvms
\begin{equation}
\label{eqq98}
 N^n \,p^n=c_1 X^2+c_2 XY+c_3 Y^2
 \end{equation}

\sli
\noindent with $\,(c_1,\,c_2,\,c_3) \in \R^3\,$ and $\,{c_2}^2-4 c_1 c_3 
<0\,$ (the form is definite positive) 

\svpl
$\bullet$ Let us suppose that for each $\,N\in \CF$, there exist at most 
two primes $\,p\in\CP\backslash\CF\,$ satisfying this relation, and let 
$\,\Delta\in\N\,$ be the product of all those primes $\,p$\, : for all 
$\,p\in\CP\backslash\CF$\, such that $\,\ord_p(\Delta)=0$, for all 
$\,N\in\CF$, for all $\,\alpha \in R^{pr}(N^2 p^2)\cup R^{pr}(N p)$, 
we have $\,\alpha\cdot x_1\neq x_1$. We deduce from proposition 
\ref{prop15666} that
 
\ssvms\sli
\begin{equation}
\label{eqq100}
\begin{array}{|l}\forall \,p\in \CP\backslash\CF \;\mbox{ such that }\; 
\ord_p(2ab D D' \Delta)=0 \; \mbox{ and } \left(\frac{a}{p}\right)=-1 
\\ \forall \,\alpha\in R(p)\cup R^{pr}(p^2)\qquad \,\forall \,i\in\{\,1 
\dots l\,\} \qquad\alpha\cdot x_1 \neq x_i
\end{array}\end{equation}

\ssvpl
Now assume that for some $\,N\in \CF,$ equation (\ref{eqq98}) is solvable 
for at least three distinct primes $\,p_1,\,p_2,\,p_3$. We have three 
points $\,(X_i : Y_i)_{i=1,2,3} \in \PP^1(\Q)\,$ such that

\ssvms
\begin{equation}
\label{eqq100b}
\forall\,i=1\dots 3\qquad N^n\,{p_i}^n=c_1 {X_i}^2+c_2 X_i Y_i+c_3 {Y_i}^2
\end{equation}

\sli
\noindent If $\,(X_i : Y_i)=(X_j : Y_j)\,$ for $\,i\neq j$, then $\,p_j
(X_i , Y_i)=\pm p_i (X_j , Y_j)\,$ whence $\,\alpha_i \in p_i R$\, since 
$p_i\wedge p_j=1$, a contradiction. Thus $(X_1 : Y_1)$, $(X_2 : Y_2)\,$ 
and $\,(X_3 : Y_3)\,$ are three distinct points of $\,\PP^1(\Q)$. By 
relation (\ref{eqq100b}) we can write

\ssvms\sli
$$\matrr{{X_1}\sli^2\sli}{X_1 Y_1}{{Y_1}^2}{{X_2}^2\sli}{X_2 Y_2}
{{Y_2}^2}{{X_3}^2}{X_3 Y_3}{{Y_3}^2}\vect{c_1\sli}{c_2\sli}{c_3} 
=N^n \vect{{p_1}^n\sli}{{p_2}^n\sli}{{p_3}^n}\in \Z^3$$

\sli
\noindent The determinant of the above matrix is $\,\prod_{i<j}(Y_j 
X_i-Y_i X_j)\in\Q^*$\, so that $\,c_1,\,c_2,\,c_3 \in \Q$\, after 
inversion of the linear system. Therefore, relation (\ref{eqq98}) becomes

\ssvms
\begin{equation}
\label{eqq99}
\kappa p^n=\alpha X^2+ \beta XY+\gamma Y^2\qquad \mbox{ for } \quad 
p\in \CP\backslash \CF  \quad\mbox{ and }\quad X,\,Y \in \Z
\end{equation}

\sli
\noindent with $\,(\kappa,\,\alpha,\,\beta,\,\gamma)\in\Z^4\,$ and 
$\,\delta=\beta^2-4\alpha\gamma<0$, which implies that $\,\delta\,$ 
is a square modulo~$p$. We deduce from proposition \ref{prop15666}
that 

\ssvms
\begin{equation}
\label{eqq184}
\begin{array}{|l}\forall \,p\in \CP\backslash\CF \;\mbox{ such that }\; 
\ord_p(2abDD')=0,\;\left(\frac{a}{p}\right)=-1\;\mbox{ and }\;\left(
\frac{\delta}{p}\right)=-1 \\ \forall \,\alpha \in R(p) \cup R^{pr}(p^2)
\quad\forall\,i \in \{1,\dots,l\}\qquad \alpha . x_1 \neq x_i
\end{array}\end{equation}

\sli
\noindent By proposition \ref{prop4}, $\,a\,$ and $\,\delta\,$ being 
strictly negative integers, there exists infinitely many primes $p
\notin\CF$ satisfying the hypothesis in relation (\ref{eqq184}) and 
(\ref{eqq100}). This leads to

\begin{prop}
\label{prop5369}
Let $x_1,\dots,x_l$ be points of $\HH^3$. There exists infinitely many 
primes $p$ such that : \;$\forall \,\alpha \in R(p) \cup R^{pr}(p^2)
\quad \forall\,i \in \{1,\dots,l\} \qquad \alpha\cdot x_1 \neq x_i$
\end{prop}

\subsection{The geodesics}\label{SepGeo}

The proof given in \cite{R-S} still holds ; we associate to any 
geodesic of $\HH^3$ a proportionality class of binary quadratic forms 
(they are complex this time). Using proposition \ref{prop3}, 
we obtain

\ssvms\sli
\begin{prop}
\label{prop11}
Let $\,L_1,\, \dots,\, L_r\,$ be geodesics of $\HH^3$. There exists 
infinitely many primes $\,p$\, such that : $\;\forall\,\alpha\in R(p)
\cup R^{pr}(p^2)\quad\forall\,i\in\{1 \dots r\}\qquad\alpha\cdot L_1 
\neq L_i$
\end{prop}

\subsection{The $\,\Gamma_{\!_R}$\,-\,closed Itgs}\label{SepItgs}

\subsubsection{The half-planes}

In this section too, we consider primes $p$\, such that 
$\ord_p(2abDD')=0$ and $\,\left(\frac{a}{p}\right)=-1$.

\begin{prop}
\label{prop12}
Let $\,\CP_1\,$ be a $\,\Gamma_{\!_R}\,$-\,closed half-plane of 
$\,\HH^3\,$ and $\,\CF\,$ a finite set of primes. There exists 
infinitely many primes $\,p\notin\CF\,$ such that

\ssvms
$$\forall\,N\in\CF\quad\forall\,\alpha\in R^{pr}(N p)\cup 
R^{pr}(N^2 p^2)\qquad\alpha\cdot \CP_1 \neq \CP_1$$
\end{prop}

\sli
\emph{Proof : }let $\,\gamma_1=\xi_1+\eta_1 \,\Omega \in \Gamma_{
\!_R}\,$ hyperbolic be such that $\,\gamma_1\cdot \CP_1 = \CP_1$.
By proposition \ref{zyva}, the trace of $\,\CP_1\,$ on $\,\C$\, is 

\ssvms\sli
\begin{equation}
\label{eqq24}
\CCD_1=\left\{\; z \in \C\shpl\Big/\shpl \Im(b\,\ol{\eta_1}\,z)=
\Im(\xi_1) \;\right\}
\end{equation}

\sli
\noindent For  $\,n=1\,$ or $\,2$, $\,p\notin\CF\,$ prime and 
$\,N\in\CF$, we take $\,\alpha=\xi+\eta\,\Omega\in R^{pr}(N^np^n)\,$ 
such that $\,\alpha\cdot \CP_1 = \CP_1$. By proposition \ref{prop6}, 
$\,\xi\,\eta\neq 0$. Then $\alpha^{-1}(\infty)=-\frac{\ol{\xi}}{
b\ol{\eta}} \in \,\CCD_1$\, which means that 

\ssvms\sli
$$-\Im\left(\cfrac{\ol{\eta_1\,\xi}}{\ol{\eta}}\right)=\Im\left(
\cfrac{\eta_1\, \xi}{\eta}\right)=\Im(\xi_1)\quad\mbox{ and }\quad
\exists\,\lambda \in \R,\;\;\cfrac{\xi}{\eta}=\cfrac{\lambda+i\,
\Im(\xi_1)}{\eta_1}$$

\sli
\noindent In fact, $\lambda \in \Q=\R\cap\F\,$ because all the 
complex numbers considered belong to the number field $\F$. Moreover 
$\,N^n p^n=\n(\alpha)=|\xi|^2-b\,|\eta|^2=|\eta|^2\,\left(\frac{
|\xi|^2 }{|\eta|^2}-b\right)=\frac{|\eta|^2}{|\eta_1|^2}\left[
\lambda^2+\Im(\xi_1)^2-b\,|\eta_1|^2\right]$. As $\,|\xi_1|^2-b\,
|\eta_1|^2=\n(\gamma_1)=1$, then $\,\Im(\xi_1)^2-b\,|\eta_1|^2=1-
\Re(\xi_1)^2\,$ so that

\ssvms\sli
\begin{equation}
\label{eqq29}
4 |\eta_1|^2 N^n p^n=|\eta|^2 \Big[4 \lambda^2+4-\tr(\gamma_1)^2\Big]
\end{equation}

\sli
\noindent By proposition \ref{prop6}, $\,\ord_p |\eta|^2=0$. Since 
$\,D'\eta_1\in\CO_{\CF}$, we have $\,{D'²}|\eta_1|² \in \Z$\, so that 
$\,\ord_p |\eta_1|^2\geq 0$. As $\,\tr(\gamma_1) \in \Z$, relation 
(\ref{eqq29}) provides $\,\ord_p(4\lambda^2)=\ord_p(4|\eta|^2\lambda^2)=
\ord_p\big[|\eta|^2(\tr(\gamma_1)^2-4)+4|\eta_1|^2 N^n p^n\big]\geq 0\,$ 
so that $\,4\lambda^2+4-\tr(\gamma_1)^2\equiv 0 \,[p]$. Therefore the 
nonnegative integer $\,c=\tr(\gamma_1)²-4\,$ is a square modulo $p$. 

\ssvpl
Assume that the integer \,$c\,$ is  a square : we set $\,\tr(\gamma_1)=
m\in\Z\,$ and $\,c=m^2-4=n^2$\, with $\,n\in\Z$, whence $\,m^2-n^2=
(|m|+|n|)\times (|m|-|n|)=4=2\times 2= 4 \times 1$. Because 
$|m|+|n| \geq |m|-|n| > 0$, we have the following alternative : 
either $\,|m|+|n|=|m|-|n|=2$\, so that $\,n=0\,$ and $\,m=
\tr(\gamma_1)=\pm 2$, or $\,|m|+|n|=4\,$ and $\,|m|-|n|=1\,$ 
so that $\,m=\tr(\gamma_1)=5/2\in \N$, a contradiction in each case. 
Hence, $\,c=\tr(\gamma_1)^2-4>0\,$ is not a square. As a consequence, 
the integers \,$c$, $\,a$\, and $\,ac\,$ are not squares in $\,\Z\,$ 
(the two last being strictly negative) so that $\,a\,$ and $\,c\,$ are 
2-independent : there exists by theorem \ref{theo3} infinitely many 
primes $\,p\notin\CF\,$ such that

\ssvms
$$\ord_p(2abDD')=0\qquad\mbox{and}\qquad\left(\frac{a}{p}
\right)=\left(\frac{c}{p}\right)=-1$$

\sli
\noindent By relation (\ref{eqq29}), these primes are the ones we 
were looking for.

\subsubsection{The half-spheres}\label{SepSph}

\begin{prop}
\label{pro8}
Let $\,\CCS=S(a_1,r)\neq S^{o}\,$ be a $\,\Gamma_{\!_R}\,$-\,closed 
itgs of $\,\HH^3\,$ and $\,\CF\,$ be a finite subset of $\,\CP$. There 
exists infinitely many primes $\,p\in\CP\backslash\CF\,$ such that

\ssvms
$$\forall\,N\in\CF\quad\forall \,\alpha\in R^{pr}(N p) \cup 
R^{pr}(N^2 p^2)\qquad \alpha\cdot \CCS \neq \CCS$$
\end{prop}

\sli
\emph{Proof : }since $\,\CCS\,$ is a $\,\Gamma_{\!_R}\,$-\,closed 
itgs of $\,\HH^3$, $\,a_1\neq 0\,$ by proposition \ref{pro7}. For 
$\,n=1\,$ or $\,2$, $\,p\in\CP\backslash\CF\,$ such that $\,\ord_p(
2abDD')=0\,$ and $\,N\in\CF$, we take $\,\alpha=\xi+\eta\,\Omega\in 
R^{pr}(N^n p^n)$\, such that $\,\alpha\cdot\CCS=\CCS=S(a_1,r)$. If 
$\,\eta=0$, then $\,\alpha\cdot (z,t)=(\xi z/\ol{\xi},t)\,$ for any 
$\,(z,t)\in\HH^3\,$ and the transformation $\,\alpha\,$ is as an 
euclidean rotation of the space $\,\R^3$. The relation $\,\alpha\cdot\CCS=
\CCS\,$ implies then $\,a_1=\alpha\cdot a_1=\xi\, a_1/\ol{\xi}\neq 0\,$ 
whence $\,\xi=\ol{\xi}\in\R\,$ and $\,\alpha=\xi=\pm N p\in pR$, 
a~contradiction. As a consequence, $\,\eta\neq 0$\, and we can proceed as 
in the proof of proposition \ref{pro7} : by proposition \ref{oboidorman},

\ssvms\sli
$$\qquad \begin{array}{c} \exists\, \eps=\pm 1 \\ \mbox{ } \\ \mbox{ } 
\end{array}\qquad\left\{\begin{array}{r@\ c@\ lcl} b\big(\ol{\eta}\,a_1
-\eps\,\eta\,\ol{a_1}\big)&=&\big(1+\eps\big) \xi&\quad &(\ref{eqq49}'') 
\sli \\ b^2 \,r^2 |\eta|^2-|\xi+b\,\eta\,\ol{a_1}|^2&=&\eps\,N^n p^n && 
(\ref{eqq48}'') \sli \\ |\xi|^2-b|\eta|^2&=& N^n p^n &&(\ref{eqq47}'')  
\end{array}\right.$$

\sli
\underline{\emph{If $\,\eps=-1$}} : we get $\,\ol{\eta}\,a_1+\eta\,
\ol{a_1}=2\,\Re(\eta \,\ol{a_1})=0\,$ and $\,\eta\ol{a_1}\in i\,\R^*$\, 
from relation (\ref{eqq49}''). We proceed as with relation (\ref{eqq104}) 
: fix $\,\eta_1\in\CO_{\F}\,$ such that $\,\eta_1\ol{a_1}\in\R^*$\, : 
then $\frac{\eta}{\eta_1}\in\R\cap\F=\Q$\, so that $\,\exists\,\lambda
\in\Q,\,\eta=\lambda\,\eta_1$. Using the relations (\ref{eqq48}'') and 
(\ref{eqq47}''), we get $\,|\xi+b\,\eta\,\ol{a_1}|^2-b^2\,r^2|\eta|^2=
|\xi|^2- b\,|\eta|^2\,$ whence $\,2\,b\,\Re(\xi\,\ol{\eta}\,a_1)+b\,
|\eta|^2\big[1+b\,(|a_1|^2-r^2)\big]=0\,$ and

\ssvms\svms
\begin{equation}
\label{eqq441}
2\,\Re\,(\xi\,\ol{\eta_1}\,a_1)=-\lambda\,q\,|\eta_1|^2
\qquad\mbox{where}\qquad q=1+b\,(|a_1|^2-r^2)
\end{equation}

\sli
Let us assume for the moment that $\,q\neq 0$. By proposition \ref{pro7}, 
$\,\zeta=\frac{a_1}{q}\in\F\,$ so that $\,2\,i\,\Im(\xi\,\ol{\eta_1}\,
\zeta)=\mu\,\sqrt{a}$\, for $\,\mu\in\Q$. Therefore, $\,2\,\xi\,\ol{\eta_1}
\,\zeta=-\lambda\,|\eta_1|^2+\mu\,\sqrt{a}\,$ and $\,4\,|\xi\,\eta_1\,
\zeta|^2= \lambda^2|\eta_1|^4-a\,\mu^2$. Injecting this in relation 
(\ref{eqq47}''), we get \,$4\,N^n \,p^n\,|\eta_1|^2|\zeta|^2 = 4\,|\xi\,
\eta_1\,\zeta|^2-4\,b\,|\eta|^2|\eta_1|^2|\zeta|^2 = -a\,\mu^2+\lambda^2
|\eta_1|^4 \big(1-4\,b\,|\zeta|^2\big)$. For integers $\,l, m, r\,$ 
relatively prime such that $\,\lambda=l/r$\, and $\,\mu=m/r \in \Q$, 
this relation becomes 

\ssvms\sli
\begin{equation}
\label{eqq45}
a\,\big(1-4\,b\,|\zeta|^2\big) (l\,|\eta_1|^2)^2=4\,a\,|\eta_1|^2
|\zeta|^2\,r^2\,N^{n}\, p^n+a^2 m^2
\end{equation}

\sli\sli
\noindent after multiplication of both sides by $\,a\,r^2$. According
to proposition \ref{pro7}, $\,1-4\,b\,|\zeta|^2\neq 0$. For $p\in\CP$\,  
such that $\,\ord_p(1-4\,b\,|\zeta|^2)=\ord_p |\zeta|^2=\ord_p 
|\eta_1|^2=0$, we deduce from relation (\ref{eqq45}) that 
$\,l\equiv 0\,[p]\,$ if and only if $\,m\equiv 0\,[p]$. If 
\,$l\equiv m\equiv 0\,[p]$, then $\,r\not\equiv 0\,[p]\,$ since 
$\,l\wedge m\wedge r=1$\, and

\ssvms\sli
\begin{equation}
\label{niou1}
(2\,r\,\ol{\eta_1}\,\zeta)\,\xi=\big(-l\,|\eta|^2+m\sqrt{a}\,\big)
\in p\CO_{\F}
\end{equation}

\noindent Moreover

\svms
\begin{equation}
\label{niou2} 
r\,\eta=l\,\eta_1\in p\CO_{\F}
\end{equation}

\noindent We have $\,D'\CI\subset D \CI_0\subset \CI$\, by 
proposition \ref{prop5} : then there exists $\,\alpha_1=
\xi_1+\eta_1\,\Omega\in\CI_0$\, such that $\,D'\alpha=D\alpha_1$. 
As $\,p\wedge D=1$, we deduce from relations (\ref{niou1}) and 
(\ref{niou2}) that $\,\alpha_1=p\,\alpha_2$, where $\,\alpha_2=
\xi_2+\eta_2\,\Omega\in\CI_0=\CO_{\F}\oplus\CO_{\F}\,\Omega$. 
Finally $\,p\wedge D'=1$\, implies that $\,xp+yD'=1$\, holds for 
two integers \,$x$, $\,y\in\Z\,$ (theorem of Bézout). We have 
then $\,\alpha=xp\alpha+y D'\alpha=xp\alpha+y D\alpha_1= p\big(x\alpha+yD\alpha_2)\in p\CI$\, since $\,D \CI_0\subset \CI$, 
a contradicition as $\,\alpha\,$ is primitive. As a consequence, 
$\,l\not\equiv 0 \,[p]\,$ and relation (\ref{eqq45}) implies that $\,a\,
\big(1-4\,b\,|\zeta|^2\big)\,$ is a square modulo $p$.

\ssvpl\sli
If $\,q=0$, we deduce from relation (\ref{eqq441}) that $\,\xi\,\ol{\eta_1}
\,a_1\in i\R$\, whence $\,\xi\in\R\,$ since $\,\ol{\eta_1}\,a_1\in i\R^*$. 
Then $\,\xi=\mu \in \Q\,$ and relation (\ref{eqq47}'') becomes $\,N^n p^n
=\mu^2-b\,|\eta_1|^2\,\lambda^2$. As in the case $\,q\neq 0$, we are lead to
 
\svms
\begin{equation}
\label{eqq451}
r^2 N^n p^n=m^2-b\,|\eta_1|^2\,l^2
\end{equation}

\sli
\noindent for integers $\,l, \,m, \,r\,$ relatively prime. Let $\,p\in
\CP\,$ such that $\,\ord_p |\eta_1|^2=0\,$ : we have $\,l\equiv 0\,[p] 
\,\Longrightarrow\,m\equiv 0\,[p]\,\Longrightarrow \alpha \in pR$, a 
contradiction. Thus $\,l\not\equiv 0\,[p]\,$ and the integer $\,b\, 
|\eta_1|^2\,$ is a square modulo $p$ by relation (\ref{eqq451}). 

\svpl
\underline{\emph{If $\,\eps=1\,$}} : relation (\ref{eqq49}'') leads to 
$\,\xi=i\,b\,\Im(\ol{\eta}\,a_1)\in i\R\,$ and $\,\xi+b\,\eta\,\ol{a_1}=
b\,\Re(\ol{\eta}\,a_1)\in\R$. Using relations (\ref{eqq48}'') and 
(\ref{eqq47}''), \ssli we get $\,N^n p^n=b^2\,{r_1}^2\, |\eta|^2 - b^2 
\,\Re^2(\eta\,\ol{a_1})=b^2 \,\Im^2(\eta\,\ol{a_1})- b\,|\eta|^2,$ so that 
$\,b^2 \,|\eta|^2 \big(|a_1|^2-{r_1}^2\big)=b\,|\eta|^2\neq 0$. Hence
\,$b\,\big(|a_1|^2-{r_1}^2\big)=1\,$ and $\,q=2$ : by proposition $\,a_1=
2\,\zeta \in \F^*$. We can set $\,\eta\,\ol{a_1}=X+Y\,\sqrt{a}\,$ with 
$\,X,\,Y \in \Q$. After multiplication of both sides by $\,|a_1|^2$, 
relation (\ref{eqq47}'') becomes 

\ssvms\sli
$$|a_1|^2\,N^n p^n=-a\,b^2\,|a_1|^2\,Y^2-b\,\big(X^2-a\,Y^2\big)
=b\,\Big[a\,\big(1-b\,|a_1|^2\big)\,Y^2-X^2\Big]$$

\sli
\noindent Let $\,p\in\CP\,$ such that $\,\ord_p |a_1|^2=0=\ord_p{b}$\, : 
as before, if $\,Y\equiv 0 \,[p] \,$ then $\, X\equiv 0\,[p] \,$ 
whence $\,\alpha\in pR$, a contradiction. Thus $\,Y\not\equiv 0\,[p]$\, 
and $\,a\,\big(1-b\,|a_1|^2\big)=a\,\big(1-4\,b\,|\zeta|^2\big)\,$ is a 
square modulo $\,p$.

\svpl
\underline{\emph{End of the proof}} : if $\,q=0$, we have $\,\eta_1\in 
\F^*\,$ and $\,\ord_b(|\eta_1|^2)\,$ is even by lemma \ref{lem6}, so that 
$\,b\,|\eta_1|^2\,$ is not a square. Therefore, there exists infinitely 
many primes $\,p\,$ such that $\,b\,|\eta_1|^2\,$ is not a square modulo 
$\,p$ (cf. section \ref{Numb}). 

\sli\sli
If $\,q\neq 0$, then $\,a\,\big(
1-4\,b\,|\zeta|^2\big)=(X^2-4)\,Y^2>0\,$ with $\,(X,Y)\in\Z\times\Q$, 
according to proposition \ref{pro7}. As we saw in the proof of proposition 
\ref{prop12}, the nonnegative integer $\,X^2-4\,$ cannot be a square. Hence, 
$\,a\,\big(1-4\,b\,|\zeta|^2\big)\,$ is not a square in \,$\Q$, and there 
exists once again infinitely many primes $\,p\,$ such that$\,a\,\big(1-4\,b
\,|\zeta|^2\big)\,$ is not a~square modulo $\,p$.

\sli\sli
In each case, after the exclusion of the prime factors of a finite set of 
rational numbers, we still have infinitely many primes $\,p\,$ for which : 

\ssvms
$$\forall\,N\in\CF\quad\forall\,\alpha\in R^{pr}(N^n p^n)
\qquad \alpha\cdot \CCS\neq\CCS$$

\subsubsection{Synthesis}

Let $\,\CCS_1\neq S^{o}, \dots,\, \CCS_l\,$ be $\,\Gamma_{\!_R}$\,-\,closed 
itgs of \,$\HH^3$, to which we apply proposition \ref{prop15666}. We~know 
by propositions \ref{prop12} and \ref{pro8} that there are infinitely many 
primes $\,p\in\CP\backslash\CF\,$ such that :
$\,\forall \,N\in\CF,\;\forall\, \alpha=\xi+\eta\,\Omega \in 
R^{pr}(N p)\cup R^{pr}(N^n p^n),\; \alpha\cdot \CCS_1 \neq \CCS_1$\, 
whence, by proposition \ref{prop15666} 

\ssvms\sli
\begin{prop}
\label{prop20}
Let $\,\CCS_1\neq S^{o},\dots,\CCS_l\,$ be $\,\Gamma_{\!_R}\,$-\,closed 
itgs of $\,\HH^3$. There exists infinitely many primes $\,p \in \CP\,$ 
such that

\ssvms\sli
$$\forall \,\alpha \in R(p) \cup R^{pr}(p^2)\quad\forall \,i\in\{\,1
\dots l\,\}\qquad\alpha\cdot \CCS_1 \neq \CCS_i$$
\end{prop}

\subsection{Conclusion}\label{neuf}

\underline{\emph{Proof of proposition \ref{prop15}}} : let $\,X_{_R}\,$ 
be a manifold of class $(K_2^S)$, $\,\Lambda\subset X_{_R}\,$ a non-empty 
set that is not of type $(S^{o })$ such that \index{Lambda@$\Lambda$}
\,$\Lambda\subset z_1 \,\cup\,\dots\,\cup\,z_l\,\cup\,L_1\,\cup\,\dots
\,\cup\,L_r\,\cup\,\Sigma_1\,\cup\,\dots\,\cup\,\Sigma_s\,$ where the 
$\,z_i\,$ are points, the $\,L_j\,$ are closed geodesics and the 
$\,\Sigma_k\,$ are $\,\Gamma_{\!_R}$\,-\,closed itgs  of $X_{_R}$. 
We assume moreover that area$(\Lambda)\neq 0$, unless $\,\Lambda$\, is 
contained in a finite union of isolated points and closed geodesics. 
We look for a modular correspondence $\CC_p$\index{CC@$\CC,\,\CC_p$} 
separating $\Lambda$ ; to this end, we shall adapt the method of 
\cite{R-S} \S 2.4.

\ssvpl
$\centerdot$ \emph{$\Lambda$ is finite} : using proposition \ref{prop5369},
we finish the proof as in \cite{R-S}.

\ssvpl
$\centerdot$ \emph{$\Lambda$ is infinite and contained in a finite union 
of closed geodesics} : using proposition \ref{prop11}, we finish the proof 
as in \cite{R-S}.

\ssvpl
$\centerdot$ \emph{$\Lambda$ is contained in a finite union of 
$\,\Gamma_{\!_R}$\,-\,closed itgs and area($\Lambda) \neq 0$} : since it 
is not of type $(S^o)$, we may write $\,\Lambda\subset\Sigma_1\cup\,\dots
\,\cup\Sigma_s\,$ where the $\,\Sigma_k\,$ are $\,\Gamma_{\!_R}$\,-\,closed 
itgs, $\,\Sigma_1 \neq S^o\,$ and area($\Sigma_1 \cap \Lambda)\neq 0$. 
Let $\,\CCS_1, \dots,\,\CCS_s\,$ be liftings of these itgs $\,\Sigma_k\,$ 
in $\,\HH^3$. Then $\,\CCS_1 \neq S^{o}$\, and there exists by proposition 
\ref{prop20} a prime $p$ such that : $\;\forall\,\alpha\in R(p)\cup R^{pr}(p^2),
\;\forall\,k=1\dots r,\;\alpha\cdot \CCS_1\neq \CCS_k$. Therefore $\,\CC_p(
\Sigma_1)\,$ and $\,\CC_{p^2}(\Sigma_1)\,$ consist of $\,\Gamma_{\!_R}$\,-\,closed 
itgs all distinct from the~$\Sigma_k$. By lemma \ref{llem666a}, the sets 
\,$\mu_1=\CC_p(\Sigma_1)\cap(\Sigma_1 \cup\,\dots\,\cup\Sigma_l)\,$ and 
$\,\mu_2=\CC_{p^2}(\Sigma_1)\cap(\Sigma_1\cup\,\dots\,\cup\Sigma_l)$\, have 
zero area. The same goes for \,$\nu_1=\left\{\,z\in X_{_R}\,\big/\,\CC_p(z)\cap 
\mu_1\neq\emptyset\,\right\}\,$ and $\,\nu_2=\left\{\,z\in X_{_R}\,
\big/\,\CC_{p^2}(z)\cap\mu_2\neq\emptyset\,\right\}$, so that there 
exists $\,z\in\Lambda\cap\Sigma_1\backslash (\nu_1\cup\nu_2)$. We finish 
the proof as in \cite{R-S}.

\svpl
\noindent\underline{\emph{Proof of theorem \ref{tryst2}}} : let 
$\,\Lambda \subset X_{_R}\,$ be a set satisfying the statement 
of proposition \ref{prop15}. There is a modular correspondence $\,\CC\,$ 
separating $\,\Lambda$. Let $\nu$ be an arithmetic quantum limit on 
$\,X_{_R}\,$ : it is associated to a sequence of eigenfunctions of 
the Laplacian $\,\Delta\,$ and the Hecke operators $\,\left(T_n\right)_{n\in\N}$, 
hence of $\,T=T_C$. Using proposition \ref{pro1}, we deduce that $\ssup{\nu}
\neq\Lambda$, which proves theorem \ref{tryst2}.\index{CC@$\CC,\,\CC_p$}
\index{TC@$T_{_C},\,T_p$}

\np
\appendix
\addtocontents{toc}{\string\setcounter{tocdepth}{1}}

\section{Of closed itgs in $\,X_{_R}$}\label{stgiferm}
\index{closed@closed}\index{itgs@$\mbox{itgs}$}

Let $\,\CCS\neq S^{o}\,$ be a itgs of $\,\HH^3\,$ closed for 
$\,\Gamma_{\!_R}$ : by proposition \ref{prop7}, there exists a 
compact subset $\,\CF\subset\CCS\,$ and a group $\,\Gamma_0\subset
\Gamma_{\!_R}\,$ such that $\,\CCS=\Gamma_0 \cdot \CF,$ \emph{i.e.} 

\ssvms
\begin{equation}
\label{app1}
\forall\,x\in\CCS\quad \exists\,\gamma \in \Gamma_0\qquad \gamma
\cdot x \in \CF\sli
\end{equation}

\noindent We set $\,t_0=\mbox{inf}\left\{\, t \,\big/\,(z,t)\in \CF\,\right\}
=\mbox{min}\left\{\, t \,\big/\,(z,t) \in \CF\,\right\}>0\,$ because $\,\CF\,$ 
is compact.

\subsection{The half-spheres}

We shall write $\,\CCS=S(a_1,r)$\, with $\,a_1 \in \C^*\,$ (cf. 
proposition \ref{pro7}) and $\,r>0$. For $\,0 < t < t_0$, we fix
$\,x=(z,t)=\left(a_1\left[1+\frac{\sqrt{r^2-t^2}}{|a_1|}\right],t
\right)\in\CCS\backslash\CF$. Let $\,\gamma=\xi+\eta\,\Omega\in 
\Gamma_0$. If $\,\eta \neq 0$, $\,\gamma\,$ is hyperbolic and 
$\,\ol{\eta} a_1 \in i\R^*$\, (cf. proof of proposition \ref{pro7}). 
Moreover 

\ssvms\sli
\begin{equation}
\label{app1bis}
\gamma\cdot x=(\tilde{z},\tilde{t})=\left(\,*\,,\,\cfrac{t}{|\xi
+b\eta\ol{z}|^2+ b^2 |\eta|^2 t^2}\,\right)
\end{equation}

\sli
\noindent by relation (\ref{kwa25}). As $\,\eta\ol{z} \in i\R$, we 
have \,$|\xi+b\eta\ol{z}|^2\geq \Re(\xi)^2 > 1\,$ by hyperbolicity 
of $\,\gamma$ : hence $\,\tilde{t} < t < t_0\,$ and $\,\gamma\cdot x
=(\tilde{z},\tilde{t})\notin \CF$. If $\,\eta=0$, then $\,\gamma=
\pm I_2\,$ and $\,\gamma\cdot x=x \notin \CF$. Therefore $\,\gamma
\cdot x \notin \CF$\, for all $\,\gamma\in\Gamma_0$, a contradiction 
with relation (\ref{app1}).

\subsection{The half-planes}

We shall write $\,\CCS=\CCD\oplus\R_+^*\, \textbf{j}$. Let 
$\,\gamma_0=\xi_0+\eta_0\,\Omega \in \Gamma_0\,$ with $\,\eta_0\neq 0$ : 
by relation (\ref{eqq102}) we have \,$\CCD=\left\{\; z\in \C \;\Big/\; 
\Im(b\,\ol{\eta_0}\,z)=\Im(\xi_0) \;\right\}=\left(\frac{-\ol{\xi_0}}{
b\ol{\eta_0}}\,;\,\frac{\xi_0}{b\ol{\eta_0}}\right)$. For $\,\lambda
\in\R$, we set $\,z_{\lambda}=\frac{-\ol{\xi_0}}{b\ol{\eta_0}}+
\frac{\lambda}{b\ol{\eta_0}} \in \CCD$. Let $\,\gamma=\xi+\eta\,\Omega
\in\Gamma_0$. If $\,\eta\neq 0,$

\ssvms\sli
\begin{equation}
\label{app2}
\xi+b\eta\ol{z_{\lambda}}=\xi-\cfrac{\xi_0 \eta}{\eta_0}+
\cfrac{\lambda\eta}{\eta_0}=\eta \left[ \cfrac{\xi}{\eta}-
\cfrac{\xi_0}{\eta_0}+\cfrac{\lambda}{\eta_0}\right]
\end{equation}

\sli
\noindent As $\,\gamma\in \CI$, then $\,D'^2 |\eta|^2 \in \N\,$ whence 
$\,D'^2 |\eta|^2 \geq 1$. As \,$\n(\gamma)=|\xi|^2-b|\eta|^2=1$, we have 
$\,\left|\frac{\xi}{\eta}\right|^2=b+\frac{1}{|\eta^2|}\leq b+D'^2$\, and

\ssvms
\begin{equation}
\label{app3}
\forall \,\gamma\in \Gamma_{\!_R}\;\mbox{ with }\;\eta\neq 0\qquad 
\left|\frac{\xi}{\eta}-\frac{\xi_0}{\eta_0}\right|^2\leq 4\left(b+
D'^2\right)
\end{equation}

\sli
\noindent We fix $\,\lambda > \big[2\sqrt{\left(b+ D'^2\right)}+D'\big]
|\eta_0|$, $\,t\in]0,t_0[\,$ and $x=(z_{\lambda},t)\in\CCS\backslash\CF$. 
From relations (\ref{app3}) and (\ref{app2}), we deduce that  
$\,|\xi+b\eta\ol{z_{\lambda}}|^2 > D'^2 |\eta|^2 > 1$, whence $\,\gamma
\cdot x\notin \CF\,$ by relation (\ref{app1bis}). If $\,\eta=0$, then 
$\,\gamma=\pm I_2$\, and $\,\gamma\cdot x= x\notin \CF$. Thus we get the 
same contradiction with relation (\ref{app1}) as before. 

\subsection{Synthesis}

We have shown that any $(K_2^S)$\,-\,manifold $\,X_{_R}$\, contains 
at most one closed itgs, the~projection of $\,S^{o}=S(0,1/\sqrt{b})\,$ 
in $\,X_{_R}$.\index{S0@$S^{o}$} Now, $\,\Gamma_{\!_R}$\, acts on the 
hyperbolic surface $\,S^{o}\,$ -- which is equivalent to $\,\HH^2$ -- 
as a subgroup of its orientation preserving isometries ; we proceed 
exactly the same way as in \cite{Eich} \S 4 to prove that $\,\Gamma_{\!_R}
\backslash S^{o}\,$ is compact for a maximal order  $\,R$, hence for 
any order (cf. if $\,R\subset R'$\, maximal order, then $\,\left[\Gamma_{\!_R}
\,:\,\Gamma_{\!_{R'}}\right]<\infty$).

\svpl
Set $\,\CI=\Z \,[i_1,i_2,i_3,i_4]$\, and $\,\CJ=\R\,[i_1,i_2,i_3,i_4]= 
\CI \otimes \R=\CA \otimes \R$. The images of the sets $\, \CI(1)$\, and 
$\,\CJ(1)$\, under the mapping $\,\varphi\,$\index{ffi@$\varphi$} defined 
in relation (\ref{kwa27}) are $\,\Gamma_{\!_R}$\, and 

\ssvms\sli
\begin{equation}
\label{app7}
G_{\!_R}=\left\{\,\matr{\xi}{\eta}{b\ol{\eta}}{\ol{\xi}} \in 
\mbox{M}(2,\C) \;\Big/\; |\xi|^2-b\,|\eta|^2=1\,\right\}
\end{equation}

\sli
\noindent the group of isometries induced by $\,\CJ=\CA\otimes\R$. 
We already know from relation (\ref{eqqbienbourrable}) that $\,\gamma
\cdot\left(\,0,\,1/\sqrt{b}\right)\in S^{o}$\, for all $\,\gamma\in 
\Gamma_{\!_R}$. Furthermore :

\ssvms
\begin{llem}
\label{oboi2}
The mapping \,$\psi\,:\,\gamma\longmapsto\gamma\cdot\left(\,0,\,1/
\sqrt{b}\right)\,$ defines a continuous surjection from $\,G_{\!_R}$\, 
to $\,S^{o}$. Set $\,\CCM_c=\left\{\,\xi+\eta\,\Omega \in G_{\!_R}\;\big/
\;|\xi|\leq c,\;|\eta|\leq c\,\right\}\,$ : for every $\,c>0$, the set 
$\,\psi(\CCM_c)\,$ is compact. 
\end{llem}

\ssvpl
\emph{Proof : }let $\,\gamma=\xi+\eta\,\Omega\in G_{\!_R}$ ; if $\,\eta
\neq 0$, we have
 
\svms
$$\gamma\cdot\vvect{0\svpl}{1/\sqrt{b}}=\vvect{\cfrac{\xi}{ 
b\ol{\eta}}-\cfrac{1}{ b\ol{\eta}}\,\cfrac{\xi}{|\xi|^2+b|\eta|^2}\sli}
{\cfrac{1}{\sqrt{b}\,\left(|\xi|^2+b|\eta|^2\right)}}=\vvect{ 
\cfrac{\xi}{b\ol{\eta}}\left(1-\cfrac{1}{1+2\,b\,|\eta|^2}
\right)\sli}{\cfrac{1}{\sqrt{b}\,\left(1+2\,b\,|\eta|^2\right)}}$$

\sli
\noindent since $\,|\xi|^2=1+b\,|\eta|²$, so that 

\ssvms\sli
$$\forall\,\gamma\in G_{\!_R}\qquad \psi(\gamma)=\left(\,
\cfrac{2\,\xi\,\eta}{1+2\,b\,|\eta|^2},\,\cfrac{1}{\sqrt{b}
\,\left(1+2\,b\,|\eta|^2\right)}\right)\in S^{o}$$

\sli
\noindent If $\eta=0$, then \,$\psi(\gamma)=(0,1/\sqrt{b})$\, as 
$\,\gamma=\pm I_2$, a particular case of the above expression. We get 
this way all the points of $\,S^{o}\,$ when $\,\left(\,|\eta|,
\,\arg{\,\xi\eta}\,\right)\,$ runs $\,[0,+\infty[ \times ]-\pi,\pi]$\, 
and $\,\psi\,$ is a surjection. From its expression, $\,\psi\,$ is moreover 
continuous on $\,G_{\!_R}$\, : as $\,\CCM_c\,$ is a compact subset of 
M$(2,\C)$\, for all $\,c>0$, then $\,\psi(\CCM_c)\,$ is compact. We can
also note that from the definition of $\,\psi$, we have

\ssvms
$$\forall\,\gamma_1, \,\gamma_2 \in G_{\!_R}\quad \psi(\gamma_1.\gamma_2)=
\gamma_1\cdot \psi(\gamma_2)$$ 

\ssvpl
Let us set $\,M_c=\left\{\,\alpha=x_1 i_1+x_2 i_2+x_3 i_3+x_4 i_4\in 
\CJ=\CA \otimes\R \;\big/\; \forall\,j=1\dots 4,\, |x_j| \leq c \right\}$
$\cap\left\{\,\alpha\in\CJ \;:\; \n(\alpha)=1\, \right\}\,$ for $\,c>0$. 
We have (see \cite{Eich} \S 4)

\begin{llem}
\label{proprarien}
Let $\,\CA\,$ be an indefinite division algebra over $\,\Q$\, and 
$\,\CI=\Z\,[i_1,i_2,i_3,i_4]$ a maximal order of $\,\CA$. There exists 
$\,c>0\,$ fixed such that 

\ssvms
$$\forall\,\alpha \in \CA\otimes\R\;\mbox{ with }\;\n(\alpha)=1
\quad \exists\,\eps\in \CI(1) \qquad \eps\alpha\in M_c$$
\end{llem}

\sli
\noindent After application of the mapping $\,\varphi$, the above 
relation becomes

\ssvms\sli
$$\exists\,c'=c\sqrt{1-a}>0\quad\forall\,\gamma\in G_{\!_R}\quad 
\exists\,\gamma_0 \in \Gamma_{\!_R} \qquad \gamma_0\gamma\in \CCM_{c'}$$

\sli
\noindent so that : $\;\exists\,c'>0,\;\forall\,\gamma\in G_{\!_R},\; 
\exists\,\gamma_0\in\Gamma_{\!_R},\;\gamma_0\cdot \psi(\gamma)=\psi(
\gamma_0.\gamma) \in \psi(\CCM_{c'})$. By lemma \ref{oboi2}, we know 
that : \;$\forall\,x\in S^{o},\;\exists\,\gamma\in G_{\!_R},\;x=
\psi(\gamma)$. We deduce that the set $\,\CF=\psi(\CCM_{c'})\,$ is a 
compact subset of $\,S^{o}\,$ such that

\ssvms
$$\forall\,x\in S^{o}\quad \exists\,\gamma_0\in \Gamma_{\!_R}\qquad 
\gamma_0\cdot x \in \CF $$

\sli
\noindent which means that $\,S^{o}=\Gamma_{\!_R}\cdot\CF$. As a consequence, 
the itgs $\,S^{o}\,$ is closed for $\,\Gamma_{\!_R}$, whence

\begin{prop}
Let $\,X_{_R}$\, be a $\,(K_2^S)$\,-\,manifold. The half-sphere 
$\,S^{o}=S(0,1/\sqrt{b})$\, is the only itgs of $\,\HH^3$ closed for 
$\,\Gamma_{\!_R}$ ; its projection in $\,X_{_R}\,$ is then compact.
\end{prop}

\subsection{Complement : the case $\,a>0$}

 To see if the choice of $\,a<0\,$ is the cause of 
the~lack of closed itgs in the $\,(K_2^S)$\,-\,manifolds, we define 
as in section \ref{zebuth} a class \,$(K_1^S)$\, of quotient manifolds 
$\,X_{_R}\,$ by taking\index{K2S@class $(K_1^S)$, class $(K_2^S)$}

\ssvms
\begin{equation}
\label{app4}
 a \in \N\qquad b \in \CP\qquad \left(\cfrac{a}{b}\right)=-1\qquad
\left(\cfrac{-1}{b}\right)=1\qquad\left(\cfrac{-3}{b}\right)=1 
\end{equation}

\noindent For example, \,$a=2$\, and $\,b=13\,$ are appropriate ; 
actually, we just have to take the opposite of $\,a\,$ to~get from 
the $(K_2^S)$\,-\,manifolds to the $(K_1^S)$ ones. The properties 
seen in section~\ref{QuotSpace} still hold, except that the conjugation 
in $\,\F\,$ does not coincide with the complex conjugation anymore. In 
the definition of the $\,\Gamma_{\!_R}$\,-\,closed itgs, we impose moreover 
that the considered hyperbolic elements $\,\gamma=\xi+\eta\,\Omega\,$ 
satisfy $\,\eta\neq 0$. This time, the itgs $\,P^{o}=\R\oplus\R_+^*\, 
\textbf{j}\simeq \HH^2$ is left invariant under the action of all the 
isometries induced by $\,\CA \otimes \R$\, \emph{i.e.} the group 
SL$(2,\R)$\, since $\,\sqrt{a} \in \R$. The image of $\,(x,t)\in 
\R\times\R_+^*\simeq P^{o}$\, under the action of 

\svms
$$\gamma=\matr{a}{b}{c}{d} \in \mbox{SL}(2,\R)$$

\noindent is given by 

\svms
$$\begin{array}{rccc|lcl} \gamma \cdot 
\left(\begin{array}{c} x \svpl \\ t  \end{array}\right) 
&=& \left(\begin{array}{c}\tilde{x}\svpl\\ \tilde{t}\end{array}\right)
\in P^{o} &=& \left(\begin{array}{c} \cfrac{a x+b}{d} \ssvpl\sli \\ 
\cfrac{a\,t}{d} \end{array}\right) &&\mbox{ if $\,c=0$} \\ &&&& \\
&&&& \left(\begin{array}{c} \cfrac{a}{c}-\cfrac{1}{c} \, 
\cfrac{c x+d}{(c x+d)^2+c^2 t^2} \ssvpl\sli \\ \cfrac{t}{(c x 
+ d)^2+c^2 t ^2}\end{array}\right)&&\mbox{ if $\,c\neq 0$}
\end{array} $$

\ssvpl
\noindent If $\,c=0$, we have 

\svms
$$\tilde{x}+i \tilde{t}=\cfrac{a(x+it)+b}{d}= \gamma(x+it)$$

\sli
\noindent If $\,c\neq 0$, we get 

\ssvms
$$\begin{array}{r@\ c@\ l} \tilde{x}+i \tilde{t} & = & \cfrac{a}{c}-
\cfrac{c(x-it)+d}{c\,\left|c(x+it)+d\right|^2}\; 
= \; \cfrac{a}{c}-\cfrac{1}{c\,\left[c(x+it)+d\right]} \ssvpl 
\\ & = & \cfrac{a\,\left[c(x+it)+d\right]-1}{c\,\left[c(x+it)+d\right]} 
\; = \; \cfrac{ac\,(x+it)+bc}{c\,\left[c(x+it)+d\right]} \quad 
\mbox{since}\quad ad-1=bc \ssvpl \\ & = & \cfrac{a\,(x+it)+b}{c
(x+it)+d} \; = \; \gamma(x+it)\end{array}$$

\sli
\noindent and we recognize in each case the fractional linear action of 
SL$(2,\R)$ on $\,\HH^2$. Therefore the discrete group $\,\Gamma_{\!_R}
\subset$ SL$(2,\R)$ has the same action on $\,P^{o}\,$ and on $\,\HH^2$ : 
we deduce that the quotient $\,\Gamma_{\!_R}\backslash 
P^{o}\,$ is compact and that the itgs $\,P^{o}\,$ is closed for 
$\,\Gamma_{\!_R}$\, for any order $\,R$\, (cf. \cite{Eich} \S 4). 

\svpl
 Let us verify that it is the only one in $\,\HH^3$. For $\,x=(z,t)\in\HH^3$, 
we define $\,f(x)\overset{def}{=}\frac{\Im(z)}{t}$. 
Let $\,\gamma=\xi+\eta\,\Omega\in\Gamma_{\!_R}$ : we have

\ssvms
$$\begin{array}{rccc|lcl} \gamma \cdot x &=& \left( \begin{array}{c} 
\tilde{z} \ssvpl \\ \tilde{t} \end{array} \right) & = &
\left(\begin{array}{c} \xi^2 z \ssvpl \\ \xi^2  t\end{array}\right) 
&& \mbox{if }\; \eta=0 \\ &&&&&& \ssvms \\ &&&& \left(\begin{array}{c} 
\cfrac{\xi}{b\olf{\eta}}-\cfrac{1}{b\olf{\eta}} 
\,\cfrac{\olf{\xi}+b\olf{\eta} \ol{z}}{
|\olf{\xi}+b\olf{\eta} z|^2+b (\olf{\eta})^2 t^2} \ssvpl\sli 
\\  \cfrac{ t}{|\olf{\xi}+b\olf{\eta} z|^2+b (\olf{\eta})^2 t^2}
\end{array}\right) && \mbox{if }\; \eta\neq 0\end{array}$$

\sli
\noindent Since $\,\xi,\;\eta \in \F \subset \R$, we easily compute 
in each case that $\,f(\gamma\cdot x)=f(x)$\, for all $\,\gamma\in 
\Gamma_{\!_R}$\, and $\,x\in\HH^3$. Let the itgs $\,\CCS\neq P^{o}\,$ 
be closed for $\,\Gamma_{\!_R}$ : there exists a compact subset 
$\,\CF\subset\CCS\,$ and a group $\,\Gamma_0\subset\Gamma_{\!_R}\,$ 
such that $\,\CCS=\Gamma_0\cdot\CF\,$ \emph{i.e.} 

\ssvms
$$\forall\,x\in\CCS\qquad \exists\,\gamma \in \Gamma_0\qquad 
\gamma\cdot x \in \CF$$

\sli
\noindent Since the function $\,f\,$ is continous on $\,\HH^3$\, and 
invariant under $\,\Gamma_{\!_R}$, this function is bounded on the 
compact set $\,\CF$, hence on $\,\CCS$\, according to previous relation. 
We can now get to the desired contradiction :

\begin{itemize}

\item[.] If $\,\CCS\,$ is a half-plane, its trace is a line $\,\CCD\neq\R$\, 
since $\,\CCS\neq P^{o}$. Fix $\,z_0 \in \CCD\backslash\R$\, and set
$\,x_t=(z_0,t)\in\CCS\,$ for all $\,t>0$. Then $\,|f(x_t)|\longrightarrow 
\infty\,$ as $\,t \rightarrow 0$, a~contradiction with relation.

\ssvms\sli
\item[.] If $\,\CCS=S(a,r)\,$ is a half-sphere, we have $\,a+ir \notin 
\R\,$ or $\,a-ir \notin \R\,$ since $\,r\neq 0$\, : we may assume without 
loss of generality that $\,a+ir \notin \R$. Set $\,x_t=(a+i\sqrt{r^2-t^2}
,t)\,$ for $\,t \in ]0,r]$ : then $\,f(x_t) \sim \left|\Im(a)+r
\right|/t \longrightarrow \infty\,$ as $\,t\rightarrow 0$, a contradiction 
again. Hence :
\end{itemize}

\begin{prop}
Let $\,X_{_R}\,$ be a $\,(K_1^S)$\,-\,manifold. The half-plane $\,P^{o}
=\R\oplus\R_+^* \textbf{j}$\, is the only itgs of $\,\HH^3$ closed for 
$\,\Gamma_{\!_R}$ ; its projection in $\,X_{_R}\,$ is thus compact.
\end{prop}

\sli
Therefore, the lack of closed itgs is not specific to the 
$(K_2^S)$\,-\,manifolds, as we could have thought \emph{a priori}. 
Moreover, we may verify using relation (\ref{eqq24}) that $\,P^{o}$\, 
is the only $\,\Gamma_{\!_R}$\,-\,closed half-plane in any 
$(K_1^S)$\,-\,manifold, that we also have restrictions on the
closed geodesics (they link two points of $\,\PP^1(\R)$) and on the 
$\,\Gamma_{\!_R}$\,-\,closed half-spheres, which justifies \emph{a 
posteriori} our choice to deal with the \,$(K_2^S)$\,-\,manifolds rather 
than with the \,$(K_1^S)$\,-\,ones.

\section{Of $\,\Gamma_{\!_R}$\,-\,closed itgs}\label{stgigammaferm}
\index{GammaR@$\Gamma_{_R}$!$\Gamma_{_R}$\,-\,closed}

\subsection{Families of $\,\Gamma_{\!_R}$\,-\,closed half-planes of 
$\,\HH³$}

Take a hyperbolic element $\,\gamma=\xi+\eta\,\Omega \in \Gamma_{\!_R}\,$ :  
$\,z_1=\gamma^{-1}(\infty)\,$ and $\,z_2=\gamma(\infty)\,$ are distinct 
points of $\,\C$\, since $\,\tr(\gamma) \neq 0$. Let us define the itgs 
$\,\CP_{\gamma}=\CCD_{\gamma} \oplus \R^*_+ \textbf{j}\,$ where 

\ssvms
\begin{equation}
\label{eqq103b}
\CCD_{\gamma}=(z_1,z_2)=\left\{\; z\in \C \;\Big/\; \Im(b\,\ol{\eta}\,z)=
\Im(\xi)\;\right\}
\end{equation}

\sli
\noindent Since $\,\tr(\gamma) \in \Z \subset \R$, we have $\,\gamma(\CCD_{
\gamma})=\CCD_{\gamma}\,$ and $\,\gamma\cdot\CP_{\gamma}=\CP_{\gamma}\,$ 
according to Proposition \ref{zyva} : the half-plane $\,\CP_{\gamma}\,$ 
is a $\,\Gamma_{\!_R}$\,-\,closed itgs of $\,\HH^3$. Let us fix $\,(t,\,u) 
\in \Z²$\, and look for an element in $\,\Gamma_{\!_R}$\, of the form 

\ssvms
$$\gamma=\gamma_{t,u}=\underbrace{x+yu \sqrt{a}}_{\xi}+\underbrace{y
(1+t\sqrt{a})}_{\eta}\,\Omega\,$$

\sli
\noindent with $\,x,\,y\in\Z^*$. As $\,x\neq 0$, $\,\gamma_{t,u}$\, is 
hyperbolic and 

\ssvms
\begin{equation}
\label{app12}
\n(\gamma_{t,u})=1 \;\Longleftrightarrow \; x^2-\big[\underbrace{ a u^2 + 
b(1-a t^2)}_{d}\big] y^2=1
\end{equation} 

\sli
\noindent By Fermat's Theorem on the Equation of Pell, we can solve this 
equation for non-trivial integers $\,x,\,y$\, as soon as $\,d=au²+b\,(1-
a t^2)\in\N\,$ is not a~square in $\,\Z$. Let us assume the contrary.

\sli\sli 
\underline{\emph{If $\,u\not\equiv 0\,[b]$\, :}}\, then $\,a\equiv du^{-2}
\,[b]$\, is a square modulo $\,b$, a contradiction.

\sli \sli
\underline{\emph{If $\,u\equiv 0\,[b]$\, :}}\, then $\,d\equiv 0\,[b]$\, 
and, as $d$ is a square, $\,d\equiv 0\,[b²]$\, since $\,b\,$ is square-free, 
whence $\,1-at^2 \equiv 0\,[b]$\, and $\,a$\, is a square modulo $\,b$, a 
contradiction again.

\sli\sli
\noindent Moreover, $\,d=b-a(b t^2-u^2)>0\,$ as soon as $\,|t|\,$ is big 
enough : in that case, we can solve the previous equation for non-trivial 
integers $\,x,\,y$. We deduce finally from relation (\ref{eqq103b}) that

\svms
$$\CCD_{\gamma_{t,u}}=\CCD(u,t)=\left(\R+\cfrac{i\,\Im(\xi)}{b |\eta|^2}
\right)\eta= \left(\R+\cfrac{u\sqrt{a}}{b(1-a t^2)}\right)(1+t\sqrt{a})$$

\sli
\noindent is the trace of a $\,\Gamma_{\!_R}$\,-\,closed itgs of $\,\HH³$, 
which proves the following Proposition :

\begin{prop}
\label{grave}
There are infinitely many $\,\Gamma_{\!_R}$\,-\,closed half-planes in 
$\,\HH^3$, \emph{e.g.} the half-planes
$\CP(t,u)=\left(\R+\frac{u\sqrt{a}}{b(1-a t^2)}\right)(1+t\sqrt{a}) 
\oplus \R_+^* \,\textbf{\emph{j}}$\, for integers $\,t\,$ and $\,u\in\Z$\, 
such that $\,d=b-a(bt^2-u^2)>0$.
\end{prop}

\ssvpl
Let us take for example $\,a=-2\,$ and $\,b=13\,$ : 

\begin{itemize}
\svms\sli
\item[.]  $\,\gamma_{0,1}=10+3 i \sqrt{2}- 3 \Omega\;$ leaves 
$\;\CP(0,1)=\left(\R+\frac{i\sqrt{2}}{13}\right) \oplus \R_+^*\, 
\textbf{j}$\; invariant.

\svms\sli
\item[.] \,$\gamma_{1,0}=25+4(1+i\sqrt{2})\,\Omega\;$ leaves  
 $\;\CP(1,0)=\R(1+i\sqrt{2})\oplus \R_+^*\, \textbf{j}$\; invariant.

\svms\sli
\item[.] \,$\gamma_{2,3}=10+3 i \sqrt{2}+(1+2 i \sqrt{2})\,\Omega\;$ leaves 
$\;\CP(2,3)=\left(\R+\frac{i\sqrt{2}}{39}\right) (1+2i\sqrt{2})\oplus \R_+^*\, 
\textbf{j}\;$ invariant.

\end{itemize}

\subsection{Families of $\,\Gamma_{\!_R}$\,-\,closed half-spheres of $\,\HH³$}

Let the half-sphere $\,\CCS=S(a_1,r)\,$ be an itgs of $\,\HH^3$\, and 
$\,\gamma=\xi+\eta\,\Omega\,$ a hyperbolic element of $\,\Gamma_{\!_R}$. We apply
 Proposition \ref{oboidorman} to the relation $\,\gamma\cdot \CCS=\CCS\,$ : 
as we saw in the proof of Proposition \ref{pro7}, we have $\,\eps=-1\,$ 
by hyperbolicity of $\,\gamma$. As a consequence,

\ssvms
$$\,\begin{array}{rcl} \gamma\cdot\CCS=\CCS & \Longleftrightarrow &
\left\{\begin{array}{r@\ c@\ l} 0 &=&\ol{\eta}\,a_1+\eta\,\ol{a_1}\sli\sli 
\\ 1&=&|\xi+b\,\eta\,\ol{a_1}|^2-b\,r^2 \,|\eta|^2 \sli\sli \\ 1&=&|\xi|^2
-b\,|\eta|^2\end{array}\right. \svpl \\  & \Longleftrightarrow & \left\{
\begin{array}{r@\ c@\ l} 0&=&\ol{\eta}\,a_1+\eta\,\ol{a_1}\sli\sli \\ 0&=&
(\xi-\ol{\xi})\,a_1+ \big[1+b\,(|a_1|^2-r^2)\big]\,\eta\sli\sli \\ 1&=&|\xi
|^2-b\,|\eta|^2\end{array}\right. \qquad(E) \end{array}$$

\sli
\noindent For $\,a_1 \in \F^*\,$ and $\,r \in \R^*\,$ such that $\,r^2
\in\Q$, the resolution of the system $\,(E)$\, leads to 

\ssvms\sli
$$\begin{array}{|r@\ c@\ l} \xi&=&X-\frac{1}{2}\left[1+b\,(|a_1|^2-r^2)
\right]\,Y \sqrt{a} \\ && \ssvms\\ \eta &=&Y \,a_1\,\sqrt{a}\end{array}$$

\sli
\noindent with $\,X,\,Y\in\Z$. The norm equation provides 

\ssvms\sli
\begin{equation}
\label{hellen}
1=X^2-\underbrace{\frac{a}{4}\left\{\left[1+b\,(|a_1|^2-r^2)\right]^2 
-4\,b\,|a_1|^2 \right\}}_{d} Y^2
\end{equation}

\noindent For $\,r^2 \in\Q\,$ close enough to $\,|a_1|^2$, we have 
$\,4\,b\,|a_1|^2 \geq \left[1+b\,(|a_1|^2-r^2)\right]^2\,$ and the 
rational number $\,d\,$ is nonnegative. Assume moreover that $\,\ord_b\, 
|a_1|^2 \geq 0\,$ and  $\,\ord_b\, r^2 \geq 0\,$ : since $\,a\equiv 4 d
\,[b]$\, is not a square modulo $\,b$, the rational number $\,d\,$ is not 
a square in $\,\Q$. Set $d=p/q$ with relatively prime integers $p$ and $q$, 
and $D=q^2 d=pq$. As the integer $D$ is not a square, the Pell equation 
$x^2-D y^2=1$ is solvable for non-trivial integers $x,\,y\in\Z$. Taking 
$X=x$ and $Y=qy$, we have non-trivial integers $\,X,\,Y\in \Z$ 
satisfying equation (\ref{hellen}), which proves the following Proposition.

\begin{prop}
\label{sph2}
Let $\,a_1\in\F=\Q[\sqrt{a}]$\, and $\,r^2\in\Q\,$ such that $\,\ord_b 
\,|a_1|^2\geq 0$, $\,\ord_b \, r^2\geq 0$ and $\,4\,b\,|a_1|^2\geq\left[1
+b\,(|a_1|^2-r^2)\right]^2$. The half-sphere $\,S(a_1,r)$\, is a 
$\,\Gamma_{\!_R}$\,-\,closed itgs of $\,\HH^3$.
\end{prop}

\ssvpl
For $\,a=-2$\, and \,$b=13$\, again :

\ssvms
\begin{itemize}

\item[.] $\,\gamma=359+168 i\sqrt{2}+18 i \sqrt{2} (2+3 i \sqrt{2}) 
\,\Omega$\; leaves $\;S\left(\frac{2}{3}+ i \sqrt{2},\sqrt{3}\right)$\, 
invariant.

\ssvms
\item[.] $\,\gamma=106133-69160 i \sqrt{2} (7+3 i \sqrt{2}) \,\Omega$\; leaves 
$\;S\left(7+3 i \sqrt{2},8\right)$\, invariant.

\ssvms
\item[.] $\,\gamma=19603-51480 i \sqrt{2} +2574 i \sqrt{2} (5+2 i \sqrt{2}) 
\,\Omega$\; leaves $\;S\left(5 + 2 i \sqrt{2},\sqrt{30}\right)$\, invariant.

\end{itemize}

\subsection{Projections on $\,X_{\!_R}$}

The Propositions \ref{grave} and \ref{sph2} have shown the existence of 
infinitely many $\,\Gamma_{\!_R}$\,-\,closed itgs in $\,\HH³$. Now, we 
have to verify that the set of their projections in $\,X_{\!_R}$\, is still 
infinite, to prove that

\begin{prop}
\label{depoulet}
There exist infinitely many $\,\Gamma_{\!_R}\,$-\,closed itgs in $\,X_{_R}$.
\end{prop}

\sli
\emph{Proof : }we shall consider here the $\,\Gamma_{\!_R}$\,-\,closed 
half-planes $\,\CP(t,0)$\, given by Proposition \ref{grave}. Let us take 
$\,t_1\,$ and $\,t_2\in \N$. We denote by $\,\CCD_1=(1+t_1\sqrt{a})\,\R\,$ 
and $\,\CCD_2=(1+t_2\sqrt{a})\,\R$\, the traces of $\,\CP(t_1,0)\,$ and 
$\,\CP(t_2,0)\,$  on $\,\C$. A circle of $\,\PP^1(\C)\,$ being entirely 
defined by three distinct points, we have for $\,\gamma=\xi+\eta\,\Omega 
\in \Gamma_{\!_R}\backslash\{\pm\Id\}$\, (so that $\,\eta\neq 0$)

\svms
$$\gamma\cdot\CP(t_1,0)=\CP(t_2,0)\,\Longleftrightarrow\,\gamma(\CCD_1)=
\CCD_2 \,\Longleftrightarrow\,\left\{\begin{array}{l} \gamma(\infty)  =  
\cfrac{\xi}{b\,\ol{\eta}} \,\in \CCD_2 \qquad (1)\ssvpl \\ \;\gamma(0)\;  
= \; \cfrac{\eta}{\ol{\xi}}\;\, \in \CCD_2 \qquad (2) \ssvpl \\ \gamma(1+
t_1\sqrt{a}) \in \CCD_2 \qquad (3)\end{array} \right.$$

\sli
\noindent The relations (1) and (2) are equivalent since $\,\eta/\ol{\xi}=
\left(b\,|\eta|^2/|\xi|^2\right) \times\xi/b\,\ol{\eta}\in \R\,\xi/b\,
\ol{\eta}$. By relation~(1), there exists $\,\lambda \in \Q\,$ such that 
$\,\xi=\lambda\,b\,\ol{\eta}\, (1+t_2\sqrt{a})$. Relation (3) provides 

\svms
$$\begin{array}{r@\ c@\ l}  \cfrac{\xi\,(1+t_1\sqrt{a})+\eta}{b\,\ol{\eta}\,
(1+t_1\sqrt{a})+\ol{\xi}} \;\in\; \CCD_2 & \Longleftrightarrow\ & \left[
\xi\,(1+t_1\sqrt{a})+\eta\right]\,\left[\xi+b\,\eta\,(1-t_1\sqrt{a})\right]
\;\in\;\CCD_2\sli\\ & \Longleftrightarrow & \xi^2 \,(1+t_1\sqrt{a})+b\,
\eta^2\,(1-t_1\sqrt{a})\;\in\;\CCD_2 \quad\mbox{since }\xi\eta\in\CCD_2
\ssvpl \\ & \Longleftrightarrow & \underbrace{\lambda^2 \,b\, (1+t_2\sqrt{a}) 
\,(1+t_1\sqrt{a})\,\ol{\eta}^2}_{z_1}+\underbrace{\left(\cfrac{1-t_1\sqrt{a}
}{1+t_2\sqrt{a}}\right)\,\eta^2}_{z_2}\;\in\;\R \end{array}$$

\ssvms\sli
\noindent The numbers $\,z_1\,$ and $\,z_2\in\F\,$ having opposite arguments, 
either they have the same module, either they are both reals. In the first 
case, we get after simplification 

\ssvms
$$b\,\lambda^2\,(1-a {t_2}^2)=1$$ 

\noindent whence $\,\ord_b(1-a 
{t_2}^2)=\ord_b |1+t_2\sqrt{a}|^2\,$ is odd, a contradiction with Lemma 
\ref{lem6}. Thus, $\,z_1\,$ and $\,z_2\,$ are real hence rational numbers 
(since $\F\cap \R=\Q$) and there exists $\,\mu\in\Q\,$ such that $\,\eta^2
=\mu\,(1+t_1\sqrt{a})\,(1+t_2\sqrt{a})$. Taking the square of the modulus, 
we get $\,|\eta|^4=\left(|\eta|^2\right)^2=\mu^2\,(1-a {t_1}^2)\,(1-a{t_2
}^2)\,$ with $\,|\eta|^2 \in \Q$, so that $\,(1-a {t_1}^2)\,(1-a {t_2}^2) 
\,$ is a square in $\,\Q$. We have hence proved that

\svms
$$\Big(\exists\,\gamma\,\in\Gamma_{\!_R},\; \gamma\cdot\CP(t_1,0)=\CP(
t_2,0)\Big) \; \Longrightarrow \; (1-a {t_1}^2)\,(1-a {t_2}^2) \mbox{ is 
a square in } \N$$

\noindent the first condition meaning that $\,\CP(t_1,0)\,$ and $\,\CP(t_2,
0)\,$ have the same projection in $\,X_{_R}$. To~fullfill the proof of the 
Proposition, we just have to find an infinite subset $\,\CCI \subset \N\,$ 
such that $\,(1-a {t_1}^2)\,(1-a {t_2}^2)\,$ is not a square in  $\,\N\,$ 
for all $\,t_1 \neq t_2 \in \CCI$, and it seems reasonable to think that 
it is possible for every negative integer $\,a$. Take $\,a=-2$\, for 
instance : we verify that all numbers between 0 and 24000 statisfy this 
relation, except from 2, 11, 12, 70, 109, 225, 408, 524, 1015, 1079, 1746, 
2378, 2765, 4120, 5859, 8030, 10681, 13860, 16647, 17615 and 21994. More 
generally, a conjectural Theorem states

\begin{conj}
 Let $\,A$, $\,B$, $\,C\,$ be integers relatively primes such that 
\,$A$\, is positive, $\,A+B\,$ and $\,C\,$ are not both even and $\,B^2-
4 AC\,$ is not a perfect square. Then there are infinitely many primes of 
the form $\,A n^2+B n +C$\, with $\,n \in \Z$.
\end{conj}

\sli
\noindent According to this Conjecture, there are for each $\,a\in\Z^-$\, 
such infinitely many primes of the form $\,1-a t^2\,$ with $\,t \in \N$\, : 
we just have to set $\,\CI=\big\{ t \in \N\,\big/\,1-at²\,\mbox{ prime}\,\big\}$\, 
to end the proof.
 
\svpl
\emph{Remark : }we used a very particular and limited subset of 
$\Gamma_{\!_R}$\,-\,closed itgs of $\HH^3$ to prove Proposition 
\ref{depoulet}. It is clear that other itgs shall provide infinitely 
many $\Gamma_{\!_R}$\,-\,closed itgs of $X_{\!_R}$.

\np

\np
\addcontentsline{toc}{section}{Index}

\printindex

\np
\addcontentsline{toc}{section}{Table of contents}
\tableofcontents


\begin{thebibliography}{1}
\addcontentsline{toc}{section}{References}

\bibitem{Ano} D.V. Anosov, Geodesic Flows on Closed Riemann Manifolds with Negative
 Curvature, 
\emph{Proceedings of the Steklov Institute of Mathematics}, \textbf{90} (1969), AMS-Providence

\bibitem{Eich}  Martin Eichler, Lectures on Modular Correspondances, 
\emph{Lectures on Mathematics and Physics}, Tata Institute of fundamental research, 1957.

\bibitem{Far} Hershel M. Farkas \& Irwin Kra, Riemann Surfaces, \emph{Graduate Text in 
Mathematics} \textbf{71}, Springer-Verlag, 1980

\bibitem{Hed} G.A. Hedlund, The Dynamics of Geodesic Flows, \emph{Bulletin of the American 
Mathematical Society}, \textbf{45} (1939) n°4, 241-260

\bibitem{Kar} H.H. Karimova, Geodesic Flows in three dimensional Spaces of Variable Spaces of 
Negative Curvature, \emph{Vestnik Moskov. Univ. Serija Mat.}, \textbf{5} (1959), 3-12

\bibitem{Lin} Lindenstrauss E. : Invariant Measures and Arithmetic 
 Quantum Unique Ergodicity. To appear in Annals of Mathematics.

\bibitem{Miya} Toshitsune Miyake, \emph{Modular Forms}, Springer-Verlag, 1989
 
\bibitem{Nar} Wladyslaw Narkiewicz, Elementary and Analytic Theory of 
Algebraic Numbers, \emph{Springer-Verlag} (1990)

\bibitem{Rat}  John G. Ratcliffe, Foundations of Hyperbolic Manifolds, 
\emph{Graduate Texts in Mathematics}, \textbf{149}, Springer-Verlag (1994)

\bibitem{R-S} Ze{\'e}v Rudnick and Peter Sarnak, The Behaviour of Eigenstates of Arithmetic 
Hyperbolic Manifolds, \emph{Communications in Mathematical Physics}, \textbf{161} (1994), 195-213

\bibitem{Sar} Peter Sarnak, Arithmetic Quantum Chaos, \emph{First R.A. Blyth Lectures}, University 
of Toronto, Preprint (1993)

\bibitem{Ser} Jean-Pierre Serre, \emph{Cours d'Arithm\'etique}, Presses Universitaires de France, 1970
 
\bibitem{Zel} Steven Zelditch, Uniform Distribution of Eigenfunctions on Compact Hyperbolic, 
\emph{Duke Math. Journal}, \textbf{55} (1987) n°4, 919-941.

\end{thebibliography}
\end{document}